\documentclass[twocolumn, tighten]{aastex63}
\usepackage{natbib}
\usepackage{multirow}

\emergencystretch=\maxdimen
\submitjournal{ApJ}
\shorttitle{Black hole parameter estimation from its shadow}
\shortauthors{Kumar \& Ghosh}
\definecolor{MyDarkBlue}{rgb}{0,0.08,0.5}
\definecolor{MyDarkRed}{rgb}{0.7,0.02,0.02}
\definecolor{MyDarkGreen}{rgb}{0.0,0.7,0.0}

\usepackage{amsmath}

\begin{document}
\title{Black Hole Parameter Estimation from Its Shadow}
%
\correspondingauthor{Rahul Kumar}
\email{rahul.phy3@gmail.com}
\author{Rahul Kumar}
\affiliation{Centre for Theoretical Physics, Jamia Millia Islamia, New Delhi 110025, India}
\author{Sushant G. Ghosh}
\affiliation{Centre for Theoretical Physics, Jamia Millia Islamia, New Delhi 110025, India}
\affiliation{Astrophysics and Cosmology Research Unit, School of Mathematics, Statistics and Computer Science, University of
	KwaZulu-Natal, Private Bag 54001, Durban 4000, South Africa}
%
\begin{abstract}
	The Event Horizon Telescope (EHT), a global submillimeter wavelength very long baseline interferometry array, unveiled event-horizon-scale images of the supermassive black hole M87* as an asymmetric bright emission ring with a diameter of $ 42 \pm 3\; \mu$as, and it is consistent with the shadow of a Kerr black hole of general relativity. A Kerr black hole is also a solution of some alternative theories of gravity, while several modified theories of gravity admit non-Kerr black holes.  While earlier estimates  for the M87* black hole mass, depending on the method used, fall in the range $ \approx 3\times  10^9 M_\odot- 7 \times 10^9 $$ M_\odot $, the EHT data indicated a mass for the M87* black hole of $(6.5 \pm 0.7) \times 10^9 M_\odot $. This offers another promising tool to estimate black hole parameters and to probe theories of gravity in its most extreme region near the event horizon. The important question arises: Is it possible by a simple technique to estimate black hole parameters from its shadow, for arbitrary models? In this paper, we present observables, expressed in terms of ordinary integrals, characterizing a haphazard shadow shape to estimate the parameters associated with black holes, and then illustrate its relevance to four different models: Kerr, Kerr$-$Newman, and two rotating regular models. Our method is robust, accurate, and consistent with the results obtained from existing formalism, and it is applicable to more general shadow shapes that may not be circular due to noisy data.
\end{abstract}


\section{Introduction}\label{sec:1}
Black holes are one of the most remarkable predictions of Einstein's theory of general relativity, which also provides a means to probe them via unstable circular photon orbits \citep{bardeen1973}. A black hole, due to its defining property at the event horizon along with the surrounding photon region, casts a dark region over the observer's celestial sky, which is known as a shadow \citep{bardeen1973,Falcke:1999pj}. Astronomical observations suggest that each galaxy hosts millions of stellar-mass black holes, and also a supermassive black hole at the nucleus of the galaxy \citep{Melia:2001dy,Shen:2005cw}. However, the majority of these black holes have very low accretion luminosity and thus are very faint. Due to relatively very large size and close proximity, the black hole candidates at the center of the Milky Way and in the nearby galaxy Messier 87, respectively, Sgr A* and M87*, are prime candidates for black hole imaging \citep{Broderick:2005xa,Doeleman:2008qh,Doeleman:2012zc}. Probing the immediate environment of black holes will not only provide images of these objects and the dynamics of nearby matter but will also enable the study of the strong gravity effects near the horizon. The Event Horizon Telescope (EHT) \footnote{\url{http://eventhorizontelescope.org/}}, a global array of millimeter and submillimeter radio observatories, is using the technique of very long baseline interferometry (VLBI) by combining several synchronized radio telescopes around the world. This Earth-sized virtual telescope has achieved an angular resolution of $20\,\mu$as, sufficient to obtain the horizon-scale image of supermassive black holes at a galaxy's center. The EHT has published the first direct image of the M87* black hole \citep{Akiyama:2019cqa,Akiyama:2019bqs,Akiyama:2019fyp,Akiyama:2019eap}. Further, fitting geometric models to the observational data and extracting feature parameters in the image domain indicates that we see emission from near the event horizon that is gravitationally lensed into a crescent shape around the photon ring \citep{Akiyama:2019fyp,Akiyama:2019eap}.\\
It turns out that photons may propagate along the unstable circular orbits due to the strong gravitational field of the black hole, and these orbits have a very important influence on quasinormal modes \citep{Cardoso:2008bp, Hod:2009td, Konoplya:2017wot}, gravitational lensing \citep{Stefanov:2010xz}, and the black hole shadow \citep{bardeen1973}.  Synge (\citeyear{Synge:1966}) and Luminet (\citeyear{Luminet:1979nyg}), in pioneering works, calculated the shadow cast by a Schwarzschild black hole, and thereafter Bardeen (\citeyear{bardeen1973}) studied the shadows of Kerr black holes over a bright background, which turn out to deviate from a perfect circle. The past decade saw more attention given to analytical investigations, observational studies, and numerical simulations of shadows \citep[see][]{Cunha:2018acu}. The shadows of modified theory black holes are smaller and more distorted when compared with the Kerr black hole shadow \citep{Bambi:2008jg,Johannsen:2010ru,Falcke:2013ola,Loeb:2013lfa,Younsi:2016azx,Giddings:2016btb,Mizuno:2018lxz,Long:2019nox,Konoplya:2019goy,Held:2019xde,Wang:2018prk,Yan:2019etp,Kumar:2019pjp,Vagnozzi:2019apd,Breton:2019arv}.\\
The no-hair theorem states that the Kerr black hole is the unique stationary vacuum solution of Einstein's field equations, however, the exact nature of astrophysical black holes has not been confirmed \citep{Johannsen:2011dh,Bambi:2017iyh}, and the possible existence of non-Kerr black holes cannot be completely ruled out \citep{Johannsen:2013rqa, Johannsen:2016uoh}. Indeed, the Kerr metric remains a solution in some modified theories of gravity \citep{Psaltis:2007cw}. For rotating black holes, significant deviations from the Kerr solution are found in modified theories \citep{Bambi:2013ufa,Berti:2015itd}. The Bardeen perspective of a shadow of a black hole in front of a planar-emitting source was applied to several black hole models, e.g., Kerr-Newman black hole \citep{Young:1976zz, de2000}, Chern-Simons modified gravity black hole \citep{Amarilla:2010zq}, Kaluza-Klein rotating dilaton black hole \citep{Amarilla:2013sj}, rotating braneworld black hole \citep{Amarilla:2011fx, Eiroa:2017uuq}, regular black holes \citep{Abdujabbarov:2016hnw, Amir:2016cen,Kumar:2019pjp}, and black holes in higher dimensions \citep{Papnoi:2014aaa,Abdujabbarov:2015rqa,Amir:2017slq,Singh:2017vfr}. The black hole shadow in asymptotically de-Sitter spacetime has also been analyzed \citep{Grenzebach:2014fha, Perlick:2018iye,Eiroa:2017uuq}. Black hole shadows have been investigated for a parameterized axisymmetric rotating black hole, which generalizes all stationary and axisymmetric black holes in any metric theory of gravity \citep{Rezzolla:2014mua,Younsi:2016azx,Konoplya:2016jvv}.\\
However, developing a methodological way to estimate parameters from astrophysical observations of a black hole image is a promising avenue to advance our understanding of black holes. The observations commonly used for the estimation of the mass and size of a black hole are based on the motion of nearby stars and spectroscopy of the radiation emitted from the surrounding matter in the Keplerian orbits, i.e., stellar dynamical and gas dynamical methods \citep{Gebhardt:2000fk,Schodel:2002vg, Shafee:2005ef}. The dynamical mass measurements from X-ray binaries only provide lower limits of the black hole's mass \citep{haring2004black,Narayan_2005,Casares:2013tpa}. Unlike for the mass, effects of the black hole's spin and any possible deviation from standard Kerr geometry are manifest only at the small radii. The two most commonly used model-dependent techniques to estimate the spin are the analysis of the K$\alpha$ iron line \citep{Fabian:1989ej} and the continuum-fitting method \citep{McClintock:2013vwa}. Though black hole parameters have been inferred in a number of contexts through the gravitational impact on the dynamics of surrounding matter \citep{matt1992iron,Narayan:2007ks,Broderick:2008sp,Steiner:2009af, Steiner:2010bt,McClintock:2011zq,  Narayan:2011eb,Bambi:2013sha}, the EHT observations can put stringent bounds on the parameters. Furthermore, it is found that the non-Kerr black hole shadows strongly depend on the deviation parameter apart from the spin \citep{Atamurotov:2013sca,Johannsen:2015qca,Wang:2017hjl, Wang:2018eui}. Thus, shadow observations of astrophysical black holes can be regarded as a potential tool to probe their departure from an exact Kerr nature, and in turn, to determine the black hole parameters \citep{Johannsen:2010ru}. Hioki and Maeda (\citeyear{Hioki:2009na}) discussed numerical estimations of Kerr black hole spin and inclination angle from the shadow observables, which was extended to an analytical estimation by Tsupko (\citeyear{Tsupko:2017rdo}). These observables, namely shadow radius and distortion parameter, were extensively used in the characterization of black holes shadows \citep{de2000, Amarilla:2010zq, Amarilla:2011fx,Amarilla:2013sj,Papnoi:2014aaa, Abdujabbarov:2015rqa, Abdujabbarov:2016hnw, Amir:2016cen, Amir:2017slq,Singh:2017vfr,Eiroa:2017uuq}. However, it was found that the distortion parameter is degenerate with respect to the spin and possible deviations from the Kerr solution; a method for discriminating the Kerr black hole from other rotating black holes using the shadow analysis is presented by Tsukamoto et al. (\citeyear{Tsukamoto:2014tja}). An analytic description of distortion parameters of the shadow has also been discussed in a coordinate-independent manner \citep{Abdujabbarov:2015xqa}. Motivated by the above, we construct observables that can uniquely characterize shadows to estimate the black hole parameters.\\
The aim of this paper is to give simple shadow observables and show their applicability to determining the black hole parameters with emphasis on the characterization of various black hole shadows of more general shape and size. The proposed observables do not presume any symmetries in the shadow and completely depend upon the geometry of the shadow. The characterization of the shadow's size and shape is not restricted to a circle and is applicable to a large variety of shadows. The prescription is applied to four models of rotating black holes to get an estimation of black hole parameters, and when compared with the existing results \citep{Tsukamoto:2014tja}, we find that our prescription gives an accurate estimation. Thus, this simple approach enables us to estimate the black hole parameters accurately, and the method is robust, as it is applicable to haphazard shadow shapes that may result from the noisy data.   \\
The paper is organized as follows. In Sect.~\ref{sect2}, we discuss the propagation of light in rotating black hole spacetime. Further, in Sect.~\ref{sect3}, we present the observables for shadow characterization and use them to estimate the parameters associated with four black holes in Sect.~ \ref{sect4}. In Sect.~\ref{sect5}, we summarize our main results. We use geometrized units $G=1$, $c=1$, unless units are specifically defined.

\section{Black Hole Shadow \label{sect2}}
\begin{figure}
	\begin{tabular}{c c}
		\includegraphics[scale=0.51]{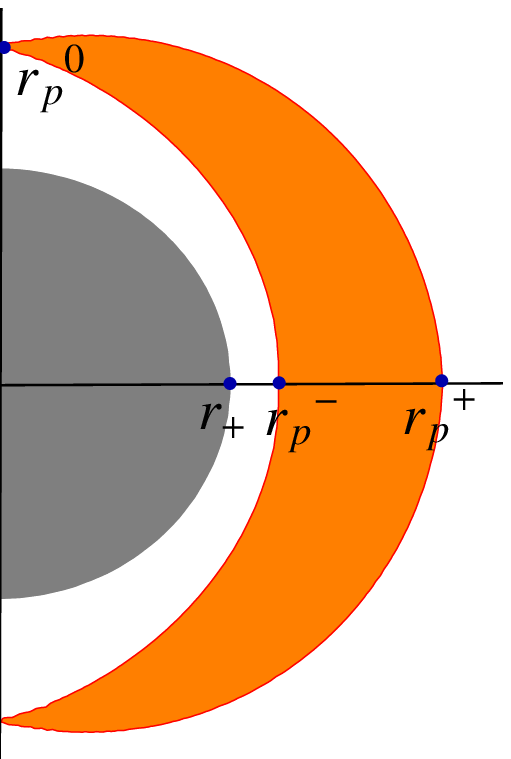}
		\includegraphics[scale=0.48]{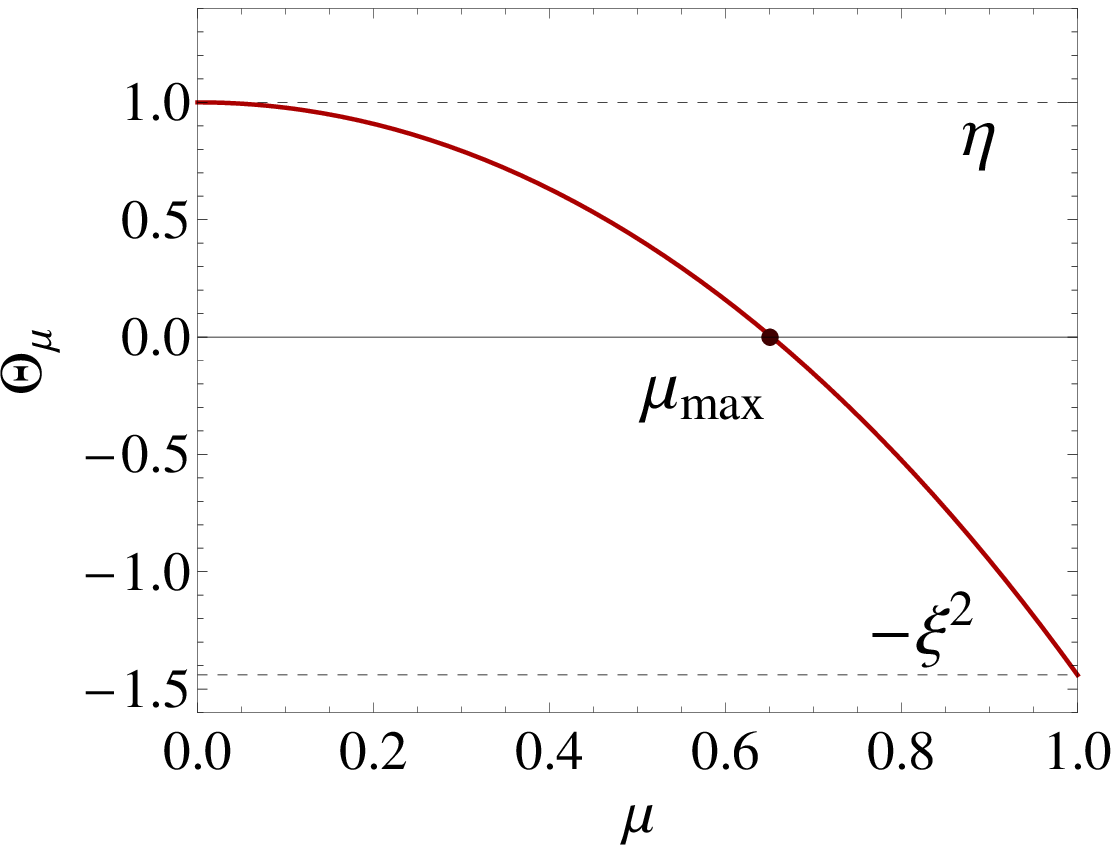} 
	\end{tabular}
	\caption{Left panel: schematic of a photon region around a rotating black hole. Right panel: variation of $\Theta_{\mu}$ with $\mu$ for $\eta=1$ and $\xi=1.2$. Horizontal dashed lines correspond to the maximum and minimum values of $\Theta_{\mu}$.}
	\label{photonregion}
\end{figure}

The metric of a general rotating, stationary, and axially symmetric black hole, in Boyer$-$Lindquist coordinates, reads \citep{Bambi:2013ufa}
\begin{eqnarray}\label{rotmetric}
ds^2 & = & - \left( 1- \frac{2m(r)r}{\Sigma} \right) dt^2  - \frac{4am(r)r}{\Sigma  } \sin^2 \theta dt \; d\phi +
\frac{\Sigma}{\Delta}dr^2  \nonumber
\\ & &+ \Sigma d \theta^2+ \left[r^2+ a^2 +
\frac{2m(r) r a^2 }{\Sigma} \sin^2 \theta
\right] \sin^2 \theta d\phi^2,
\end{eqnarray}
and
\begin{eqnarray}
\Sigma = r^2 + a^2 \cos^2\theta;\;\;\;\;\;  \Delta = r^2 + a^2 - 2m(r)r,
\end{eqnarray}
where $m(r)$ is the mass function such that $\lim_{r\to\infty}m(r)=M$ and $a$ is the spin parameter defined as $a=J/M$; $J$ and $M$ are, respectively, the angular momentum and ADM mass of a rotating black hole. Obviously metric (\ref{rotmetric}) reverts back to the \cite{Kerr:1963ud} and Kerr$-$Newman \citep{Newman:1965my} spacetimes when $m(r)=M$ and $m(r)=M-Q^2/2r$, respectively.
Photons moving in a general rotating spacetime (\ref{rotmetric}) exhibit two conserved quantities, energy $\mathcal{E}$ and angular momentum $\mathcal{L}$, associated with Killing vectors $\partial_t$ and $\partial_{\phi}$. To study the geodesics motion in spacetime (\ref{rotmetric}), we adopt the \cite{Carter:1968rr} separability prescription of the Hamilton$-$Jacobi equation. The complete set of equations of motion in the first-order differential form read \citep{Carter:1968rr,Chandrasekhar:1985kt}
\begin{eqnarray}
\Sigma \frac{dt}{d\tau}&=&\frac{r^2+a^2}{r^2-2m(r)r+a^2}\left({\cal E}(r^2+a^2)-a{\cal L}\right) \nonumber\\
 &&-a(a{\cal E}\sin^2\theta-{\mathcal {L}})\ ,\label{tuch}\\
\Sigma \frac{dr}{d\tau}&=&\pm\sqrt{\mathcal{R}(r)}\ ,\label{r}\\
\Sigma \frac{d\theta}{d\tau}&=&\pm\sqrt{\Theta(\theta)}\ ,\label{th}\\
\Sigma \frac{d\phi}{d\tau}&=&\frac{a}{r^2-2m(r)r+a^2}\left({\cal E}(r^2+a^2)-a{\cal L}\right)\nonumber\\
&&-\left(a{\cal E}-\frac{{\cal L}}{\sin^2\theta}\right)\ ,\label{phiuch}
\end{eqnarray}
with the expressions for 
$\mathcal{R} (r)$ and ${\Theta}(\theta)$, respectively, are given by 
\begin{eqnarray}\label{06}
\mathcal{R}(r)&=&\left((r^2+a^2){\cal E}-a{\cal L}\right)^2-(r^2-2m(r)r+a^2) ({\cal K}\nonumber\\
&&+(a{\cal E}-{\cal L})^2),\quad \\ 
\Theta(\theta)&=&{\cal K}-\left(\frac{{\cal L}^2}{\sin^2\theta}-a^2 {\cal E}^2\right)\cos^2\theta.\label{theta0}
\end{eqnarray}
The conserved quantity $\mathcal{Q}$ associated with the hidden symmetry of the conformal Killing tensor is related to the Carter integral of motion $\mathcal{K}$ through $\mathcal{Q}=\mathcal{K}+(a\mathcal{E}-\mathcal{L})^2$ \citep{Carter:1968rr}. One can minimize the number of parameters by defining two dimensionless impact parameters $\eta$ and $\xi$ as follows \citep{Chandrasekhar:1985kt}
\begin{equation}
\xi=\mathcal{L}/\mathcal{E},\;\;\;\;\; \eta=\mathcal{K}/\mathcal{E}^2.
\end{equation}
Due to spacetime symmetries, geodesics along $t$ and $\phi$ coordinates do not reveal nontrivial features of orbits, therefore the only concern will be mainly for Eqs. (\ref{r}) and (\ref{th}).  Rewriting Eq.~(\ref{th}) in terms of $\mu=\cos\theta$, we obtain
\begin{equation}
\Sigma \int \frac{d\mu}{\sqrt{\Theta_{\mu}}}=\int d\tau;\;\;\;\;\; \Theta_{\mu}=\eta-(\xi^2+\eta-a^2)\mu^2-a^2\mu^4. \label{theta1}
\end{equation}
Obviously $\eta\geq 0$ is required for possible $\theta$ motion, i.e., $\Theta_{\mu}\geq 0$ (see Figure \ref{photonregion}). For the Schwarzschild black hole, due to spherical symmetry, all null circular orbits are planar, i.e., orbits with  $\dot{\theta}=0$. However, in the Kerr black hole, the frame-dragging effects may lead to nonplanar orbits as well.
Indeed, planar and circular orbits around Kerr black hole are possible only in the equatorial plane ($\theta=\pi/2$) that leads to a vanishing  Carter constant ($\mathcal{K}=0$). Furthermore, generic bound orbits at a plane other than $\theta=\pi/2$ are nonplanar ($\dot{\theta}\neq 0$) and cross the equatorial plane while oscillating symmetrically about it. These orbits are identified by $\mathcal{K}>0$ (or $\eta>0$) and are commonly known as spherical orbits \citep{Chandrasekhar:1985kt}, and $\theta$ motion freezes only in the equatorial plane. Equation (\ref{theta1}) reveals that the maximum latitude of a spherical orbit, $\theta_{max}=\cos^{-1}(\mu_{max})$, depends upon the angular momentum of photons, i.e., the smaller the angular momentum of photons the larger the latitude of orbits; $\mu_{max}$ correspond to the solution of $\Theta_{\mu}(\mu)=0$. Only photons with zero angular momentum $(\xi=0)$ can reach the polar plane of the black hole ($\theta=0, \mu=1$) and cover the entire span of $\theta$ coordinates (see Figure \ref{photonregion}).\\
Depending on the values of the impact parameters $\eta$ and $\xi$, photon orbits can be classified into three categories, namely scattering orbits, unstable circular and spherical orbits, and plunging orbits. Indeed, the unstable orbits separate the plunging and scattering orbits, and their radii ($r_p$) are given by \citep{Chandrasekhar:1985kt}
\begin{equation}
\left.\mathcal{R}\right|_{(r=r_p)}=\left.\frac{\partial \mathcal{R}}{\partial r}\right|_{(r=r_p)}=0, \quad \left.\frac{\partial^2 \mathcal{R}}{\partial r^2}\right|_{(r=r_p)}\leq 0. \label{vr} 
\end{equation}

Solving Equation (\ref{vr}) yields the critical locus ($\eta_c$, $\xi_c$) associated with the unstable photon orbits, that for nonrotating black holes are at a fixed radius, e.g., $r_p=3M$ for a Schwarzschild black hole, and construct a spherical photon sphere. In the rotating black hole spacetime, photons moving in unstable circular orbits at the equatorial plane can either corotate with the black hole or counterrotate, and their radii can be identified as the real positive solutions of $\eta_c=0$ for $r$, $r_p^-$ and $r_p^+$, respectively. Photon orbit radii are an explicit function of black hole spin and lie in the range $M\leq r_p^-\leq 3M$ and $3M\leq r_p^+\leq 4M$ for the Kerr black hole, and $r_p^-\leq r_p^+$ due to the Lens$-$Thirring effect. Whereas, spherical photon orbits (orbits at $\theta\neq \pi/2$) are no longer affixed to a fixed plane and instead are three-dimensional orbits with radii in the interval $[r_p^-,r_p^+]$, i.e., for $\eta_c>0$ orbits, radii lie in the range $r_p^-<r_p<r_p^+$. Although rotating black holes generically have two distinct photon regions, viz., inside the Cauchy horizon ($r_-$) and outside the event horizon ($r_+$), for a black hole shadow we will be only focusing on the latter, i.e., for $r_p>r_+$ \citep{Grenzebach:2014fha}. The critical impact parameter $\xi_c$ is a monotonically decreasing function of $r_p$ with $\xi_c(r_p^-)>0$ and $\xi_c(r_p^+)<0$, such that at $r_p=r_p^0$ $(r_p^-<r_p^0<r_p^+)$ $\xi_c$ is vanishing. Even though for orbits at $r_p^0$ the angular momentum of photons is zero, they still cross the equatorial plane with nonzero azimuthal velocity $\dot{\phi}\neq 0$ \citep{wilkins1972bound,Chandrasekhar:1985kt}.\\
A black hole in a luminous background of stars or glowing accreting matter leads to the appearance of a dark spot on the celestial sky accounting for the photons which are unable to reach the observer, popularly known as a black hole shadow. Photons moving on  unstable orbits construct the edges of the shadow. 
A far distant observer perceives the shadow as a projection of a locus of points $\eta_c$ and $\xi_c$ on the celestial sphere on to a two-dimensional plane. Let us introduce the celestial coordinates \citep{bardeen1973}
\begin{equation}
\alpha=\lim_{r_O\rightarrow\infty}\left(-r_O^2 \sin{\theta_O}\frac{d\phi}{d{r}}\right),\quad  \beta=\lim_{r_O\rightarrow\infty}\left(r_O^2\frac{d\theta}{dr}\right).
\end{equation}
Here, we assume the observer is far away from the black hole ($r_O\to \infty$) and $\theta_O$ is the angle between the line of sight and the spin axes of black hole, namely, the inclination angle. Since the black hole spacetime is asymptotically flat, we can consider a static observer at an arbitrarily large distance, and this yields
\begin{equation}
\alpha=-\frac{\xi_c}{\sin\theta_O},\qquad \beta=\pm\sqrt{\eta_c+a^2\cos\theta_O^2-\xi_c^2\cot^2\theta_O}.\label{pt}
\end{equation} 
For an observer in the equatorial plane $\theta_O=\pi/2$, Eq.~(\ref{pt}) reduces to
\begin{equation}
\alpha=-\xi_c,\qquad \beta=\pm\sqrt{\eta_c}.\label{pt1}
\end{equation} 
Solving Eq.~(\ref{vr}) for rotating metric (\ref{rotmetric}) and using Eq.~(\ref{pt1}), the celestial coordinates of the black hole shadow boundary take the following form
\begin{eqnarray}
\alpha&=&-\frac{[a^2 - 3 r_p^2] m(r_p) + r_p [a^2 + r_p^2] [1 + m'(r_p)]}{a [m(r_p) + r_p [-1 + m'(r_p)]]}, \nonumber\\
\beta&=\pm&\frac{1}{a [m(r_p) + r_p [-1 + m'(r_p)]]}\Big[r_p^{3/2}\Big[-r_p^3(1+m'(r_p)^2)\nonumber\\
 && + m(r_p) [4 a^2 + 6 r_p^2 - 9 r_pm(r_p)] - 2r_p [2 a^2 + r_p^2 \nonumber\\
 &&- 3 r_pm(r_p)] m'(r_p)\Big]^{1/2}\Big],\label{impactparameter}
\end{eqnarray}
and whereas for $m(r)=M$, Eq.~(\ref{impactparameter}) yields
\begin{eqnarray}
\alpha &=&\frac{r_p^2 (r_p-3 M ) + a^2 (M + r_p)}{a (r_p-M)},\nonumber\\
\beta &=&\pm\frac{r_p^{3/2} (4 a^2 M - r_p (r_p-3 M )^2)^{1/2}}{a( r_p-M)},
\end{eqnarray}
and which is exactly the same as obtained for the Kerr black hole \citep{Hioki:2009na}. The contour of a nonrotating black hole shadow ($a=0$) can be delineated by
\begin{equation}
\alpha^2+\beta^2=\frac{2 r_p^4 + [m(r_p) + r_p m'(r_p)][-6 r_p^2 m(r_p) + 	2 r_p^3 m'(r_p)]}{[m(r_p) + r_p[-1 + m'(r_p)]]^2},
\end{equation}
which implies that the shadow of a nonrotating black hole is indeed a perfect circle, and further returns to $\alpha^2+\beta^2=27M^2$ for the Schwarzschild black hole $m(r)=M$. Though the shape of the shadow is determined by the properties of null geodesics, it is neither the Euclidean image of the black hole horizon nor that of its photon region, rather it is the gravitationally lensed image of the photon region. 
For instance, the horizon of Sgr A*, with $M\approx 4.3\times 10^6 M_{\odot}$ at a distance $d\approx 8.35$ kpc, spans an angular size of $20\, \mu$as, whereas its shadow has an expected angular size of $\approx 53\, \mu$as. Whereas, EHT measured the angular size of the M87* gravitational radius as $3.8\pm 0.4\, \mu$as, and its crescent-shaped emission region has an angular diameter of $42\pm 3\, \mu$as, with a  scaling factor in the range $10.7-11.5$ \citep{Akiyama:2019fyp,Akiyama:2019eap}. 

\section{ Characterization of the Shadow via New Observables \label{sect3}}
A nonrotating black hole casts a perfectly circular shadow. However, for a rotating black hole, an observer placed at a position other than in the polar directions witnesses an off-center displacement of the shadow along the direction of black hole rotation. Furthermore, for sufficiently large values of the spin parameter, a distortion appears in the shadow because of the Lense-Thirring effect \citep{Johannsen:2010ru}. Hioki and Maeda (\citeyear{Hioki:2009na}) characterized this distortion and shadow size by the two observables $\delta_s$ and $R_s$, respectively. The shadow is approximated to a circle passing through three points located at the top, bottom, and right edges of the shadow, such that $R_s$ is the radius of this circle and $\delta_s$ is the deviation of the left edge of the shadow from the circle boundary \citep{Hioki:2009na}. It was found that the applicability of these observables was limited to a specific class of shadows, demanding some symmetries in their shapes, and they may not precisely work for black holes in some modified theories of gravity \citep{Abdujabbarov:2015xqa}, which leads to the introduction of new observables \citep{Schee:2008kz,Johannsen:2015qca,Tsukamoto:2014tja,Cunha:2015yba, Abdujabbarov:2015xqa,Younsi:2016azx, Tsupko:2017rdo,  Wang:2018eui}. EHT observations can constrain the key physical parameters of the black holes, including the black hole mass and other parameters. However, EHT observations do not give any estimation of angular momentum \citep{Akiyama:2019cqa,Akiyama:2019eap}. Their measurement of the black hole mass in M87* is consistent with the prior mass measurement using stellar dynamics, but is inconsistent with the gas dynamics measurement \citep{Gebhardt:2011yw,Walsh:2013uua,Akiyama:2019eap}. Here, we would propose new observables for the characterization of the black hole shadow, which unlike previous observables \citep{Hioki:2009na}, do not require the apparent shadow shape to be approximated as a circle.\\
We consider a shadow of general shape and size to propose new observables, namely the area ($A$) enclosed by a black hole shadow, the circumference of the shadow ($C$), and the oblateness ($D$) of the shadow. The observables $A$ and $C$, respectively, are defined by 
\begin{equation}
A=2\int{\beta(r_p) d\alpha(r_p)}=2\int_{r_p^{-}}^{r_p^+}\left( \beta(r_p) \frac{d\alpha(r_p)}{dr_p}\right)dr_p,\label{Area}
\end{equation}     
and
\begin{eqnarray}
C&=&2\int\sqrt{({d\beta(r_p)}^2+{d\alpha(r_p)}^2)}\nonumber\\
&=&2\int_{r_p^{-}}^{r_p^+}\sqrt{\left(\left(\frac{d\beta(r_p)}{dr_p}\right)^2+\left(\frac{d\alpha(r_p)}{dr_p}\right)^2\right)}dr_p.\label{Circumference}
\end{eqnarray}
The prefactor 2 is due to the black hole shadow's symmetry along the $\alpha-$axis. $A$ and $C$ have dimensions of $[M]^2$ and $[M]$, respectively. A shadow silhouette can be taken as a parametric curve between celestial coordinates as a function of $r_p$ for $r_p^-\leq r_p\leq r_p^+$, i.e., a plot of $\beta(r_p)$ versus $\alpha(r_p)$. We can also characterize the shadow of rotating black hole through its oblateness \citep{Takahashi:2004xh, Grenzebach:2015oea,Tsupko:2017rdo}, the measure of distortion (circular asymmetry) in a shadow, by defining the dimensionless parameter $D$ as the ratio of horizontal and vertical diameters:
\begin{equation}
D=\frac{\alpha_r-\alpha_l}{\beta_t-\beta_b}.\label{Oblateness}
\end{equation}
The subscripts $r, l, t$, and $b$ stand for right, left, top, and bottom, respectively, of the shadow boundary. For a spherically symmetric black hole shadow, $D=1$, while for a Kerr shadow $\sqrt{3}/2\leq D< 1$ \citep{Tsupko:2017rdo}. Thus, $D\neq 1$ indicates that the shadow has distortion and hence corresponds to a rotating black hole. In particular, the quasi-Kerr black hole metric may lead to a shadow with $D>1$ or $D<1$, depending on the sign of the quadrupole deviation parameters \citep{Johannsen:2010ru}. The definitions of these observables require neither any nontrivial symmetry in shadow shape nor any primary curve to approximate the shadow. It can be expected that an observer targeting the black hole shadow through astronomical observations can measure the area, the length of the shadow boundary, and also the horizontal and vertical diameters. In what follows, we show that these observables uniquely characterize the shadow and it is possible to estimate the black hole parameters from these observables. 

The EHT observations indicated that the M87* black hole shadow is consistent with that of Kerr black hole, however, the exact nature of the Sgr A* black hole is still elusive. Astronomical observations have place constraints on their masses and distances from Earth as $M=4.3\times 10^6 M_{\odot}$ and $d=8.35$ kpc for Sgr A* \citep{Ghez:2008ms, Gillessen:2008qv,Falcke:2013ola,Reid:2014boa}, and $M=(6.5\pm 0.7)\times 10^9 M_{\odot}$ and $d=(16.8\pm 0.8)$Mpc for M87* \citep{Akiyama:2019cqa}. Presuming the exact Kerr nature of these black holes, we determine the area spanned by their shadows, the solid angle covered by them on the celestial sky, and also their angular sizes. In general, for rotating black holes, the vertical (or major $\vartheta_M$) and horizontal (or minor $\vartheta_m$) angular diameters are not the same and can be defined as 
\begin{equation}
\vartheta_M=\frac{\beta_t-\beta_b}{d},\;\;\;  \vartheta_m=\frac{\alpha_r-\alpha_l}{d},
\end{equation}
and the  solid angle is $\Omega={A}/{d^2}$. Clearly, $\vartheta_M$ is not dependent on black hole spin.
\begin{table}
	\centering	\caption{Table representing the values of observables, solid angle, and angular diameter with varying spin parameter for the Sgr A* black hole shadow.}\label{Sgr A}
	\begin{tabular}{ |p{0.6cm}|p{1.1cm}|p{1.0cm}|p{1.1cm}|p{1.37cm}|p{1.0cm}| }
		\hline
		$a/M$ &  $A$ & $C$ & $D$ &  $\Omega$ & $\vartheta_m $  \\
			 &  $(10^{20}m^2)$ & $(10^{10}m)$ & &$(10^{-3}{\mu\text{as}}^2)$ & ($\mu$as)  \\
		\hline\hline
		0.0 & 34.079 &  20.6942 & 1 &2.1818 &  52.7344 \\
		\hline
		0.10 & 34.06 & 20.6884& 0.999443& 2.18059& 52.705 \\
		\hline
		0.20 &34.0025 & 20.671& 0.997748& 2.1769& 52.6156 \\
		\hline
		0.30 & 33.9046& 20.6413& 0.994847& 2.17064& 52.4626 \\
		\hline
		0.40 &  33.7629& 20.5984& 0.990607& 2.16157& 52.239 \\
		\hline
		0.50 & 33.572& 20.5406& 0.984808& 2.14934& 51.9332 \\
		\hline
		0.60 & 33.3227& 20.4655& 0.977083& 2.13338& 51.5259 \\
		\hline
		0.70 & 32.9998& 20.3688& 0.966783& 2.11271& 50.9827 \\
		\hline
		0.80 &  32.5742& 20.2427& 0.952608& 2.08546& 50.2352 \\
		\hline
		0.90 & 31.9754& 20.0699& 0.931145& 2.04713& 49.1033 \\
		\hline
		0.998 & 30.7793& 19.776& 0.876375& 1.97055& 46.2151 \\
		\hline
	\end{tabular}
\end{table}

\begin{table}
	\centering	\caption{Table representing the values of observables, solid angle, and angular diameter with varying spin parameter for the M87* black hole shadow.}\label{M87}
	\begin{tabular}{  |p{0.6cm}|p{1.1cm}|p{1.0cm}|p{1.1cm}|p{1.37cm}|p{1.0cm}| }
		\hline
		$a/M$ &  $A$ & $C$ & $D$ &  $\Omega$ & $\vartheta_m $  \\
		&  $(10^{27}m^2)$ & $(10^{14}m)$ & &$(10^{-4}{\mu\text{as}}^2)$ & ($\mu$as) \\
		\hline\hline
		0.0 & 7.78711 &  3.12819& 1 &1.23151 &  39.6192 \\
		\hline
		0.10 &7.78278& 3.12732& 0.999443& 1.23083& 39.5971 \\
		\hline
		0.20 &7.76963& 3.12468& 0.997748& 1.22875& 39.53 \\
		\hline
		0.30 &7.74726& 3.12019& 0.994847&1.22521& 39.4151 \\
		\hline
		0.40 & 7.7149& 3.1137& 0.990607& 1.22009& 39.2471 \\
		\hline
		0.50 &7.67126& 3.10498& 0.984808& 1.21319& 39.0173 \\
		\hline
		0.60 &7.6143& 3.09362& 0.977083& 1.20418& 38.7113\\
		\hline
		0.70 &7.54052& 3.079& 0.966783& 1.19252& 38.3032 \\
		\hline
		0.80 &7.44327& 3.05994& 0.952608& 1.17714& 37.7416\\
		\hline
		0.90 &7.30644& 3.03382& 0.931145& 1.1555& 36.8912\\
		\hline
		0.998 & 6.99973& 2.65529& 0.866025& 1.10699& 34.3112 \\
		\hline
	\end{tabular}
\end{table}
Obviously, for $a=0$, $\vartheta_M=\vartheta_m= 52.7344\, \mu$as for Sgr A* and $\vartheta_M=\vartheta_m= 39.6192\, \mu$as for M87*. The shadow observables and angular diameters of the Sgr A* and M87* black hole shadows are calculated for various values of spin parameter $a$ (see Table \ref{Sgr A} and Table \ref{M87}). Nevertheless, the shadow observables for a Schwarzschild black hole take the values $A/M^2=84.823$, $C/M=32.6484$, and $D=1$, whereas for a maximally rotating Kerr black hole $A/M^2=76.6101$, $C/M=31.1998$, and $D=0.876375$.

\section{Application to Various Black Hole Spacetimes}\label{sect4}
We examine several rotating black holes such as Kerr$-$Newman, Bardeen, and nonsingular black holes. In general these black holes are given by metric (\ref{rotmetric}) with an appropriate choice of mass function $m(r)$. We assume that the observer is in the equatorial plane, i.e., the inclination angle $\theta_O=\pi/2$ for the estimation. One can use either of the two observables $A$ or $C$ along with $D$ to estimate the black hole parameters. For the sake of brevity, we shall use only $A$ and $D$ for our purpose, but shall calculate all three.

\subsection{Kerr$-$Newman black hole}
We start with a Kerr$-$Newman black hole, which encompasses Kerr, Reissner$-$Nordstrom, and Schwarzschild black holes as special cases. One can analyze null geodesics to the shadow of a Kerr$-$Newman black hole \citep{Young:1976zz,de2000}. In the case of the Kerr$-$Newman black hole, the mass function $m(r)$ has a form
\begin{equation}
m(r)=M-\frac{Q^2}{2r}.\label{Kerrnewmanmass}
\end{equation}
\begin{figure}[h!]
	\includegraphics[scale=0.7]{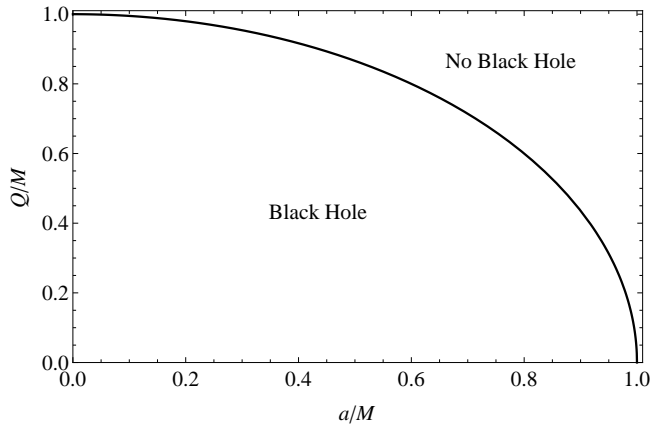}
	\caption{Allowed parametric space ($a, Q$) for the existence of a Kerr$-$Newman black hole. The solid line corresponds for the extremal black hole with degenerate horizons and demarcates the black hole case from the no black hole case.}\label{KNNoBH}
\end{figure}

\begin{figure*}
	\centering
	\begin{tabular}{c c c}
		\includegraphics[scale=0.45]{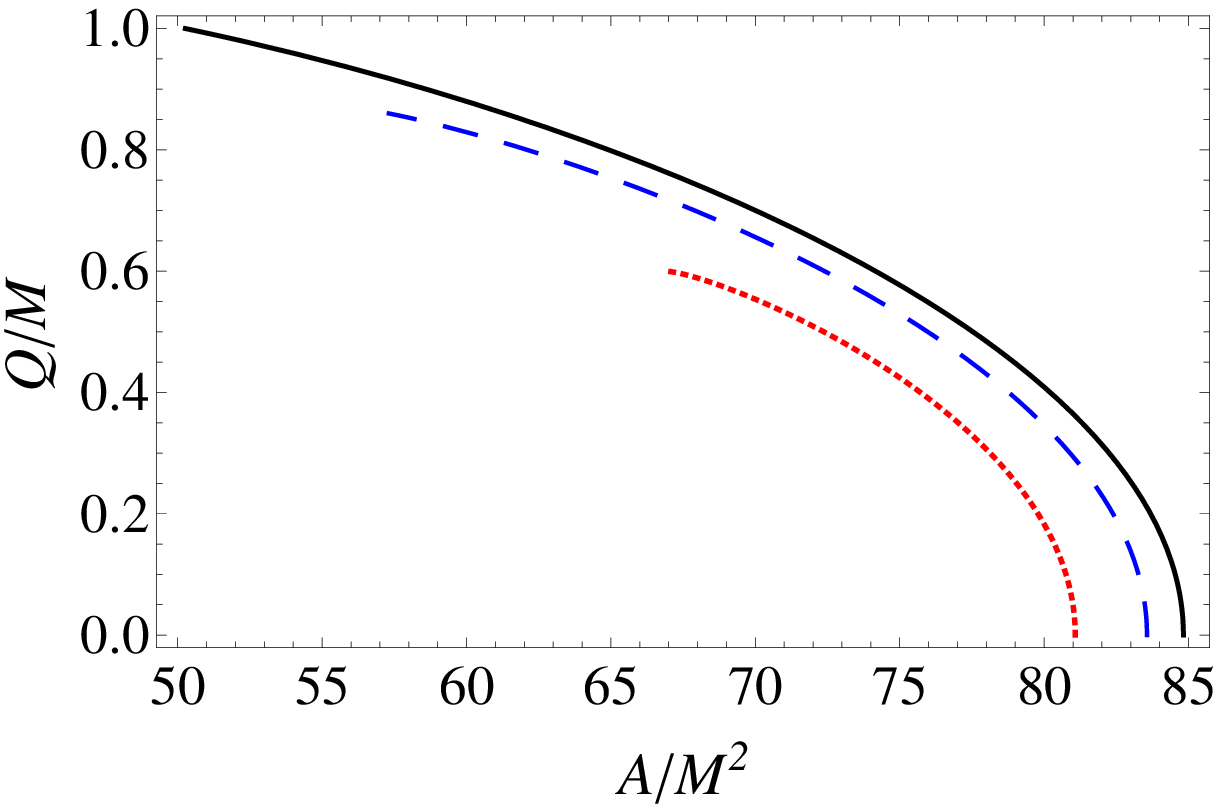}&
		\includegraphics[scale=0.45]{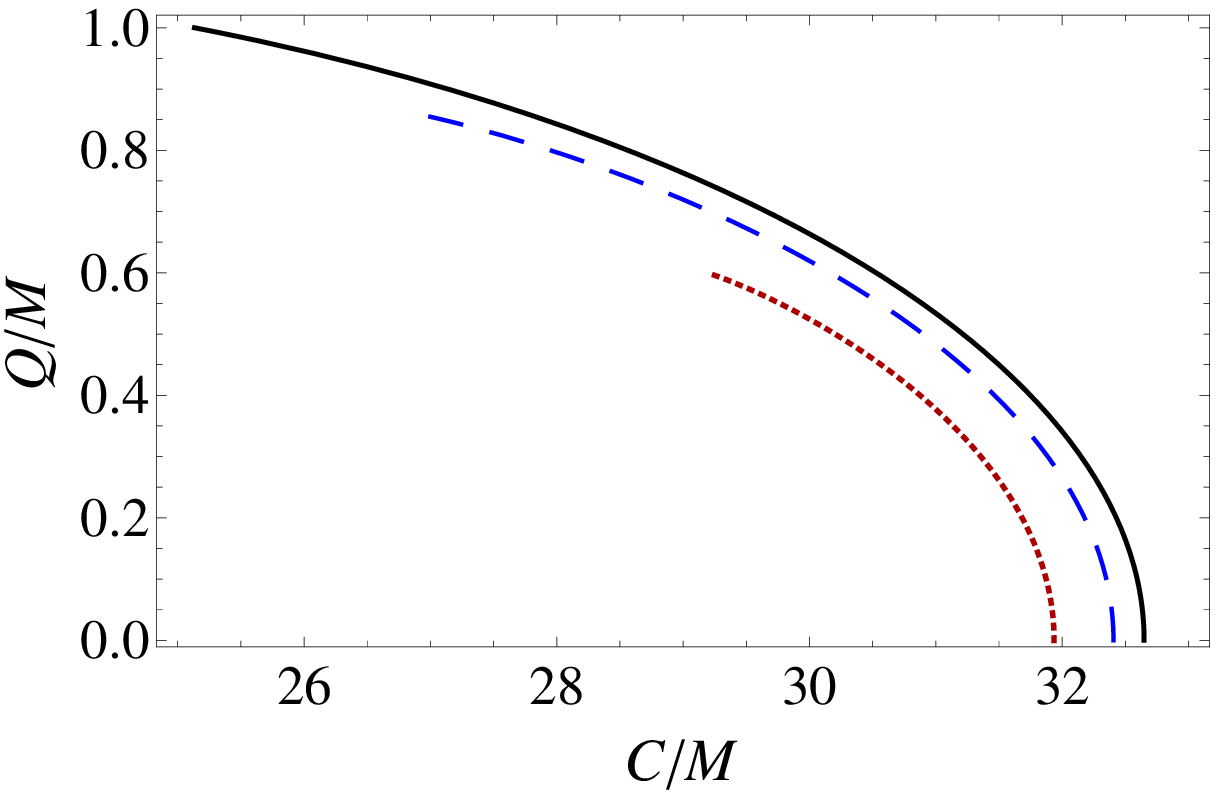}&
		\includegraphics[scale=0.45]{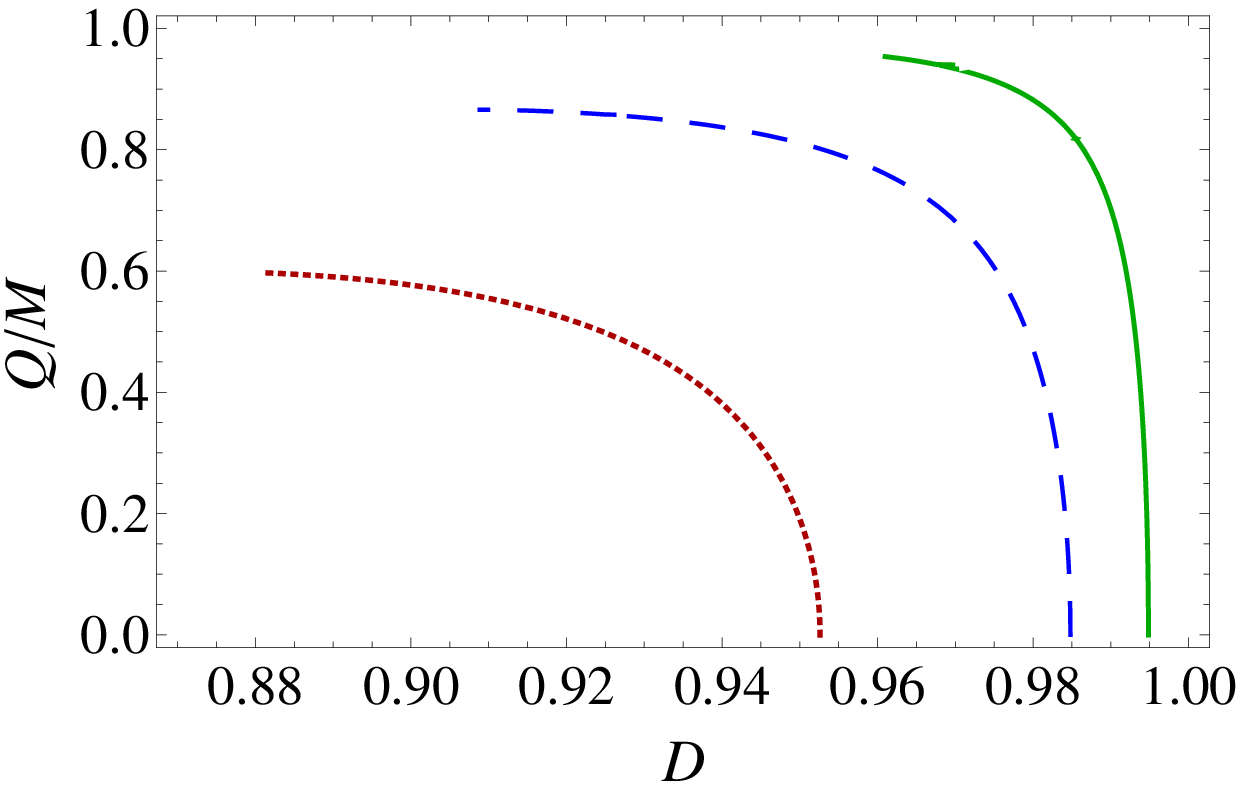} 
	\end{tabular}
	\caption{Charge parameter $Q$ vs. observables $A$, $C$, and $D$ for the Kerr$-$Newman black hole, for $a/M=0$ (solid black curve), for $a/M=0.3$ (solid green curve), for $a/M=0.5$ (dashed blue curve), and for $a/M=0.8$ (dotted red curve).}\label{KerrNewman}
\end{figure*}

\begin{figure*}
	\centering
	\begin{tabular}{c c c}
		\includegraphics[scale=0.45]{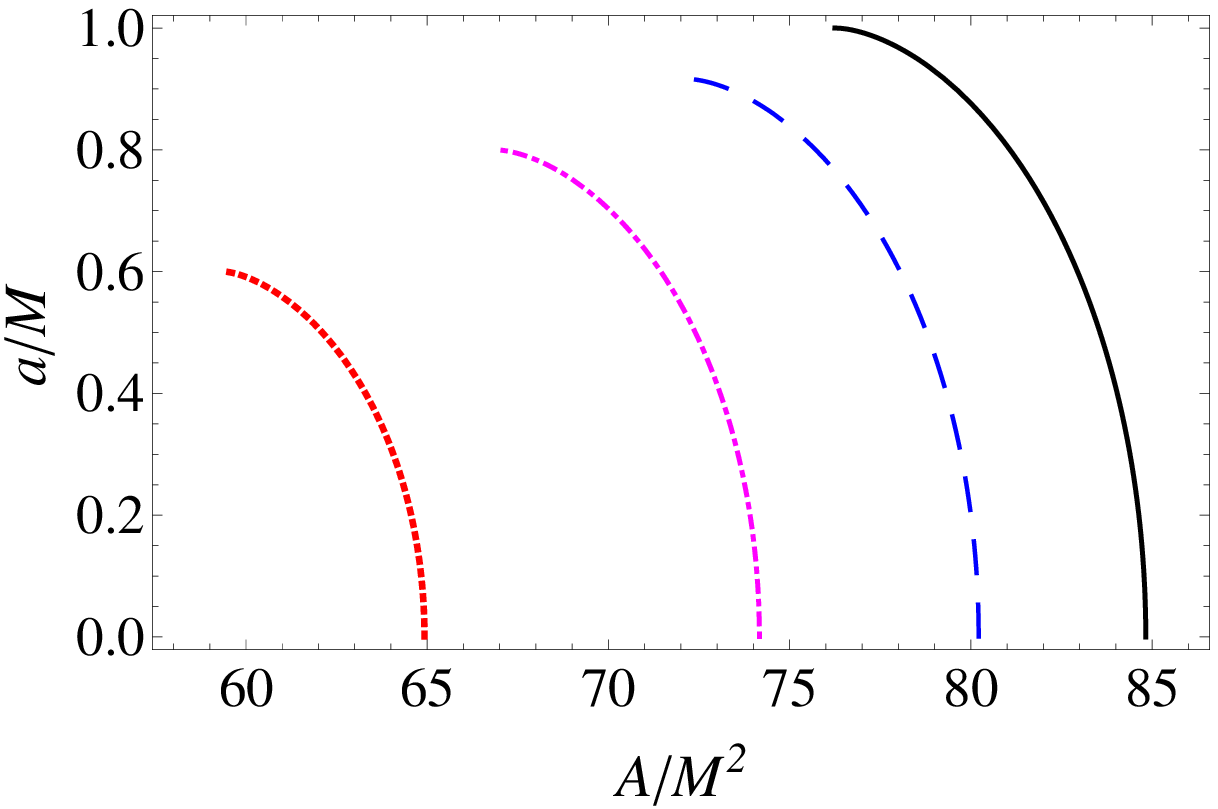}&
		\includegraphics[scale=0.45]{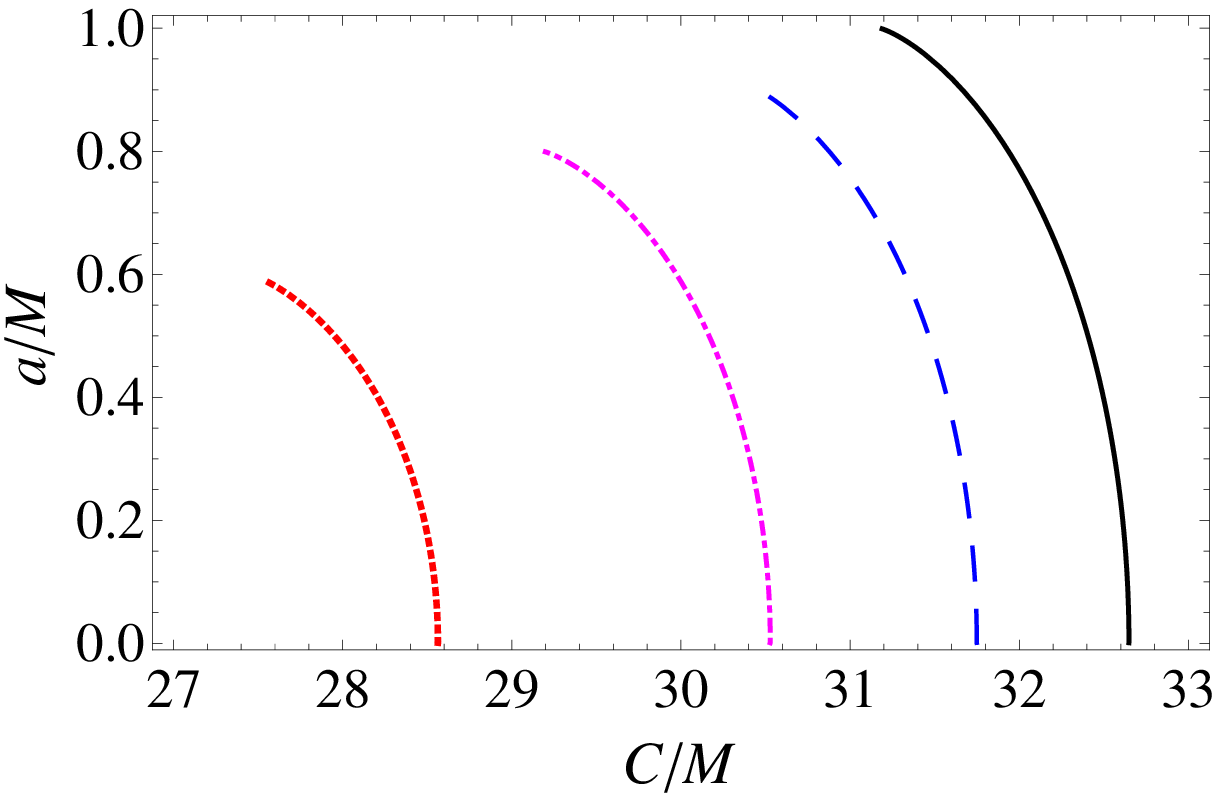}&
		\includegraphics[scale=0.45]{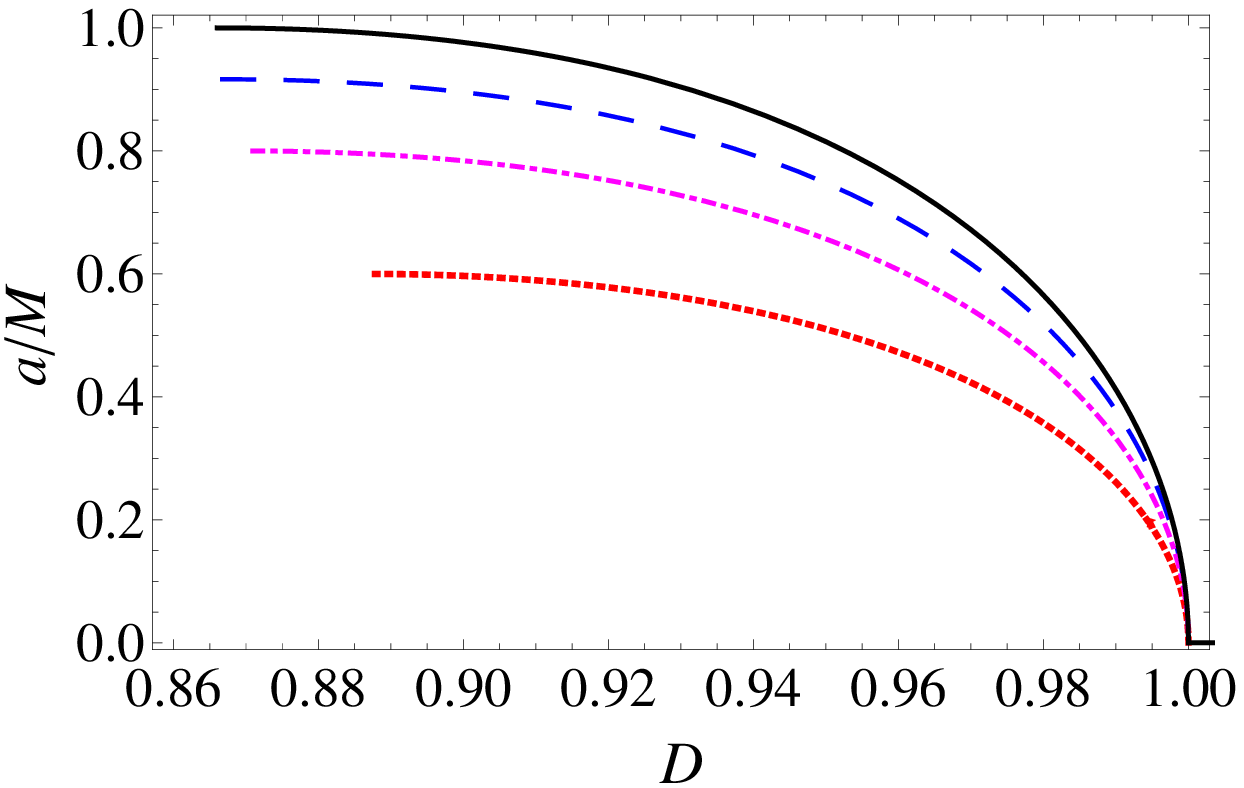} 
	\end{tabular}
	\caption{Spin parameter $a$ vs. observables $A$, $C$, and $D$ for the Kerr$-$Newman black hole, for Kerr black hole $Q/M=0.0$ (solid black curve), for $Q/M=0.4$ (dashed blue curve), for $Q/M=0.6$ (dotted dashed magenta curve), and for $Q/M=0.8$ (dotted red curve).}\label{KerrNewmanplot1}
\end{figure*}
In Figure~\ref{KNNoBH}, we have shown the allowed range of parameters $a$ and $Q$ for the existence of a black hole horizon. The Kerr$-$Newman black hole shadows are distorted from a perfect circle and possess a dent on the left side of shadow \citep{de2000}. This distortion reduces as the observer moves from the equatorial plane to the axis of black hole symmetry, and eventually disappears completely for $\theta_O=0, \pi$ \citep{de2000}. 
It is straightforward to calculate the celestial coordinates $\alpha$ and $\beta$ using the $m(r)$ in Eq.~(\ref{impactparameter}). Though for these $\alpha$ and $\beta$ the observables $A$, $C$, and $D$ could not be obtained in exact analytic form, we have calculated them approximately in the Appendix \ref{Appendix1}.

In Figures~\ref{KerrNewman} and \ref{KerrNewmanplot1}, respectively, charge $Q$ and spin parameter $a$ are plotted with varying observables $A$, $C$, and $D$. Interestingly, estimated values of black hole parameters decrease with independently increasing observables. For a far extremal black hole, parameters decrease rapidly with observables, whereas for a near extremal black hole, parameters decrease relatively slowly with increasing $D$. Therefore, one can conclude that the size of the shadow decreases with an increase in the electric charge, which is consistent with the earlier results \citep{de2000}. On the other hand, Figure~\ref{KerrNewman} suggests that shadows of Kerr$-$Newman black holes get more distorted as the charge increase. Shadow observables for Kerr$-$Newman black holes are numerically compared with those for Kerr black holes in Figure~\ref{KerrNewmanplot1}, and it can be inferred that observables for Kerr$-$Newman black holes are smaller than those for Kerr black holes for fixed values of $a$. \\
\begin{figure}
	\includegraphics[scale=0.8]{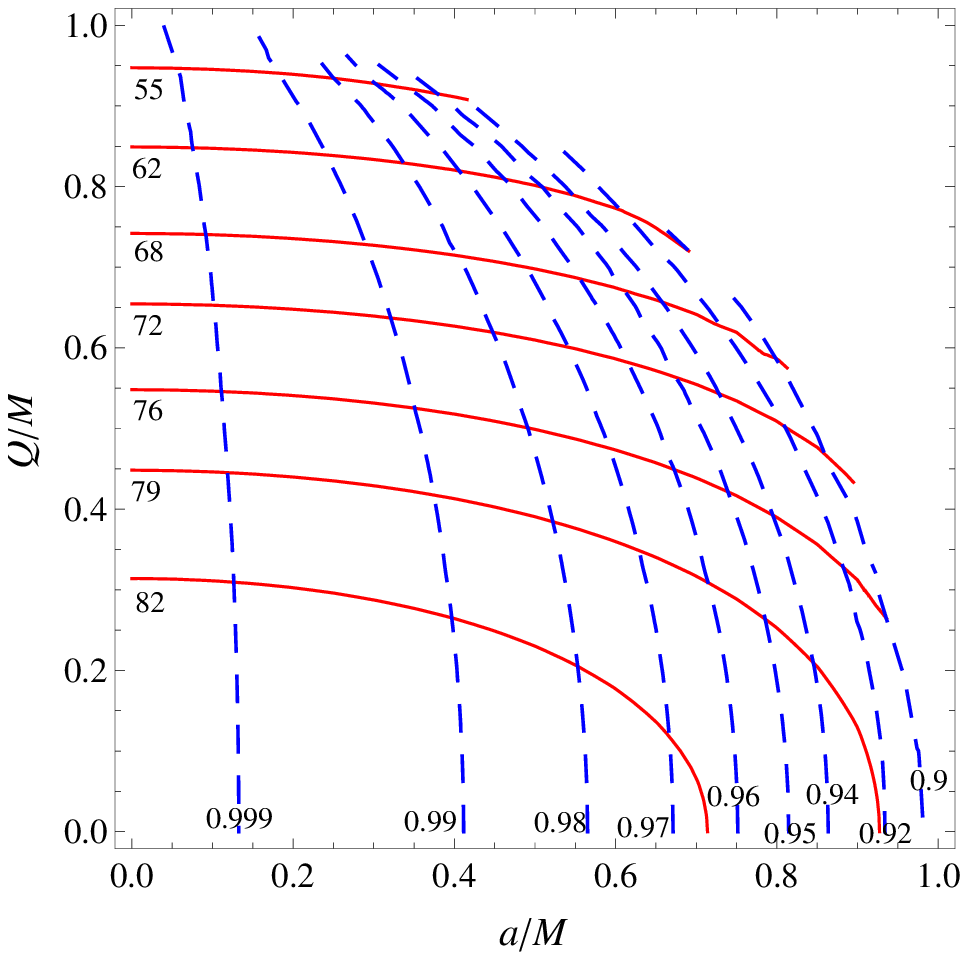} 
	\caption{Contour plot of the observables $A$ and $D$ in the plane $(a, Q)$ for a Kerr$-$Newman black hole. Each curve is labeled with the corresponding values of $A$ and  $D$. The solid red lines correspond to the area observable $A$, and the dashed blue lines correspond to the oblateness observable $D$. } \label{Kerr-Newman01}
\end{figure}

The apparent shape and size of the Kerr$-$Newman black hole shadow depend on the $a$ and $Q$ \citep{de2000}. Next, we see the possibility of estimation of $a$ and $Q$ for the Kerr$-$Newman black hole by using the two observables  $A$ and $D$, expecting that mass $M$ can be fixed through other astrophysical observations. We plot the contour map of the observables $A$ and $D$ in the ($a, Q$) plane (see Figure \ref{Kerr-Newman01}). Each point of the contour plot in Figure \ref{Kerr-Newman01} has coordinates $(a, Q)$ that can be described as a unique intersection of the lines of constant $A$ and $D$. Hence, from Figure \ref{Kerr-Newman01}, it is clear that intersection points give an exact estimation of parameters $a$ and $Q$ when one has the values of $A$ and $D$ for a Kerr$-$Newman black hole. In Table \ref{BH parameter}, we have presented the estimated values of $a$ and $Q$ for given shadow observables $A$ and $D$ for the Kerr$-$Newman black hole. \\

	\begin{table*}
	\caption{Estimated values of parameters for different black hole models from known shadow observables $A$ and $D$.}\label{BH parameter}
	\begin{tabular}{|l|l||l|l|l|l|l|l|}
		\hline
		\multicolumn{2}{|c||}{\multirow{2}{*}{Shadow Observable}} & \multicolumn{6}{c|}{Black Hole Parameters}                                                          \\ \cline{3-8} 
		\multicolumn{2}{|c||}{}                                   & \multicolumn{2}{c|}{Kerr$-$Newman} & \multicolumn{2}{c|}{Bardeen} & \multicolumn{2}{c|}{Nonsingular} \\ \hline
		$A/M^2$             & $D$                   & $a/M$            & $Q/M$             & $a/M$           & $g/M$          & $a/M$             & $k/M$             \\ \hline \hline
		82.0                               & 0.995               & 0.28284        & 0.29079         & 0.26954       & 0.28938      & 0.283082        & 0.042578        \\ \hline
		82.0                               & 0.98               & 0.55344        & 0.20523         & 0.5396       & 0.20487      &     0.553528    &  0.021183       \\ \hline
		82.0                               & 0.97                & 0.66713        & 0.11830         & 0.66130       & 0.11840      & 0.66714         & 0.0070          \\ \hline
		80.0                               & 0.995             & 0.27224          & 0.392001        & 0.24911         & 0.388453      &  0.272684        &  0.077867        \\ \hline
		80.0                               & 0.97            & 0.64252          & 0.29326         & 0.608567      & 0.292265     &    0.642787      &   0.043316       \\ \hline
		80.0                            & 0.95            & 0.80151           & 0.18123           & 0.783569       & 0.181885       &    0.801557       &  0.016468         \\ \hline
		75.0                               & 0.995             & 0.245457          & 0.56651        & 0.20174        & 0.5550      &    0.24706      &    0.165303       \\ \hline
		75.0                             & 0.95                & 0.724557        & 0.4614554        & 0.623425       & 0.45764      &     0.726799    &   0.108487      \\ \hline
		75.0                            & 0.87            & 0.9725687           & 0.231468           & N.A.          & N.A.         &  N.A.      & N.A.        \\ \hline
		70.0                            & 0.995            & 0.2180449           & 0.692545           &  0.15845          & 0.669714         &   0.2218     & 0.251415       \\ \hline
		70.0                            & 0.95            &   0.64521        & 0.620045         &  0.479736         & 0.607512         &  0.653064      & 0.199267       \\ \hline
		70.0                            & 0.90            &  0.825472          & 0.536881         &      N.A.     &   N.A.       &    0.8321695    &   0.1476488     \\ \hline
		67.0                            & 0.995            & 0.20126          &  0.75515        &   0.1337        &   0.723888       & 0.20681       &    0.30231    \\ \hline
		67.0                            & 0.98            &  0.394672         &    0.735403      &  0.263005         &   0.707514       & 0.40475       &  0.2856      \\ \hline
		67.0                            & 0.90            &  0.76314         &    0.628296      &  N.A.         &  N.A.      & 0.776511  &   0.2042426   \\ \hline
		55.0                             & 0.995            &  0.13035         &    0.944071      &  N.A.        &   N.A.       &  0.1478      &  0.497943   \\ \hline
		55.0                            & 0.95            &  0.381534         &    0.91421      &  N.A.         &  N.A.       & 0.435251       &  0.460537      \\ \hline
		55.0                            & 0.92            &  N.A.         &    N.A.      &  N.A.         &  N.A.       & 0.518359       &  0.437666      \\ \hline
		
		\end{tabular}
\end{table*}

\subsubsection{Kerr Black Hole}
When the electric charge is switched off ($Q=0$), the Kerr$-$Newman spacetime becomes Kerr with $m(r)=M$. We plot the spin parameter $a$ ($0\leq a\leq 1$) with varying observables $A$, $C$, and $D$ in Figure~\ref{KerrNewmanplot1}. It is evident that with increasing observables $A$, $C$, and $D$ the estimated Kerr spin parameter decreases. Figure~\ref{KerrNewmanplot1} indicates that the black hole shadow gets smaller and more distorted for a rapidly rotating black hole, as shown in earlier studies as well \citep{bardeen1973}.

Kerr black holes have only two parameters associated with them, namely, mass $M$ and spin $a$, however, presuming the knowledge of only mass through the stellar motion around the black hole, one has only one unknown parameter i.e., spin. The spin parameter for the Kerr black hole can be uniquely determined by knowing any one of the shadow observables defined above (see Figure~\ref{KerrNewmanplot1}).\\

\subsection{Rotating Bardeen Black Hole}
The first regular black hole was proposed by Bardeen (\citeyear{Bardeen:1968}) with horizons and no curvature singularity$-$ a modification of the Reissner$-$Nordstrom black hole.  The rotating Bardeen black hole \citep{Bambi:2013ufa} belongs to the prototype non-Kerr family with the mass function $m(r)$ given by
\begin{equation}
m(r)=M\left(\frac{r^2}{r^2 + g^2}\right)^{3/2}.
\end{equation}

\begin{figure}
	\includegraphics[scale=0.7]{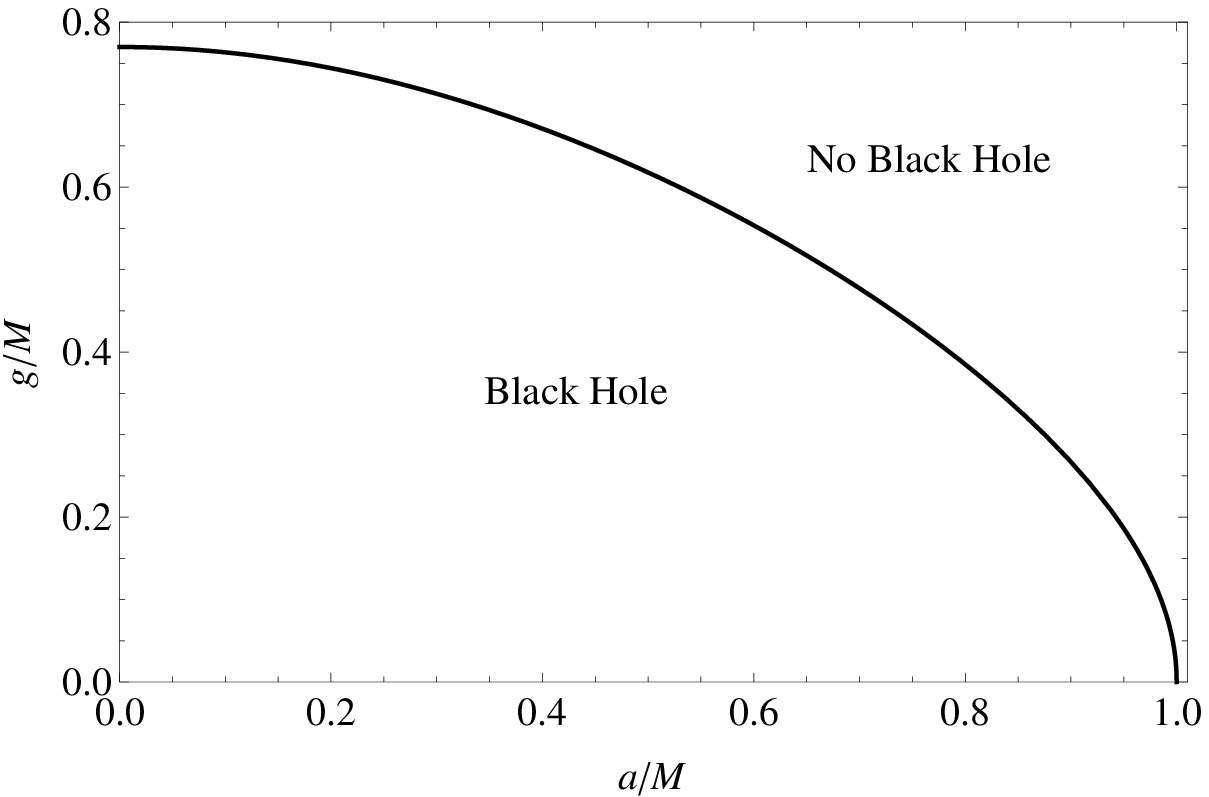}
	\caption{The allowed parametric space of $a$ and $g$ for the existence of a rotating Bardeen black hole. The solid line corresponds to the extremal black hole with degenerate horizons.}\label{BardeenNoBH}
\end{figure}

\begin{figure*}
		\begin{tabular}{c c c}
		\includegraphics[scale=0.45]{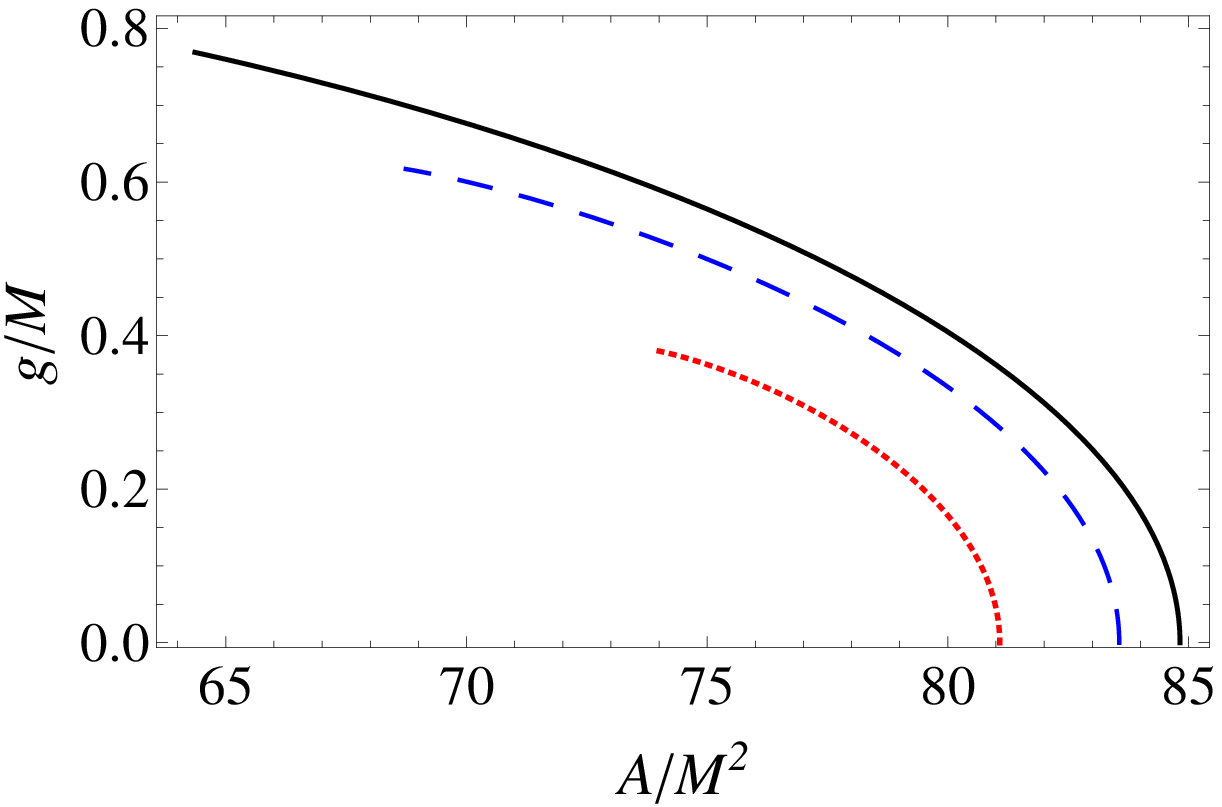}&
		\includegraphics[scale=0.45]{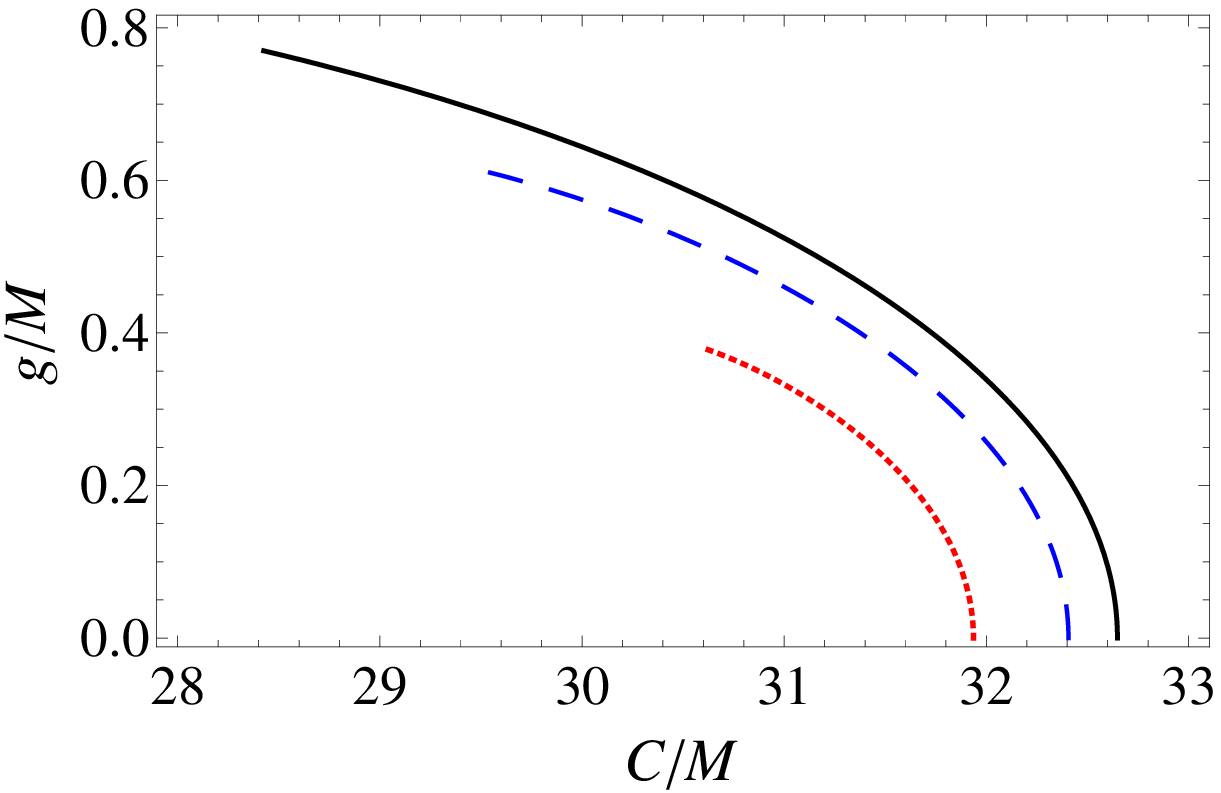}&
		\includegraphics[scale=0.45]{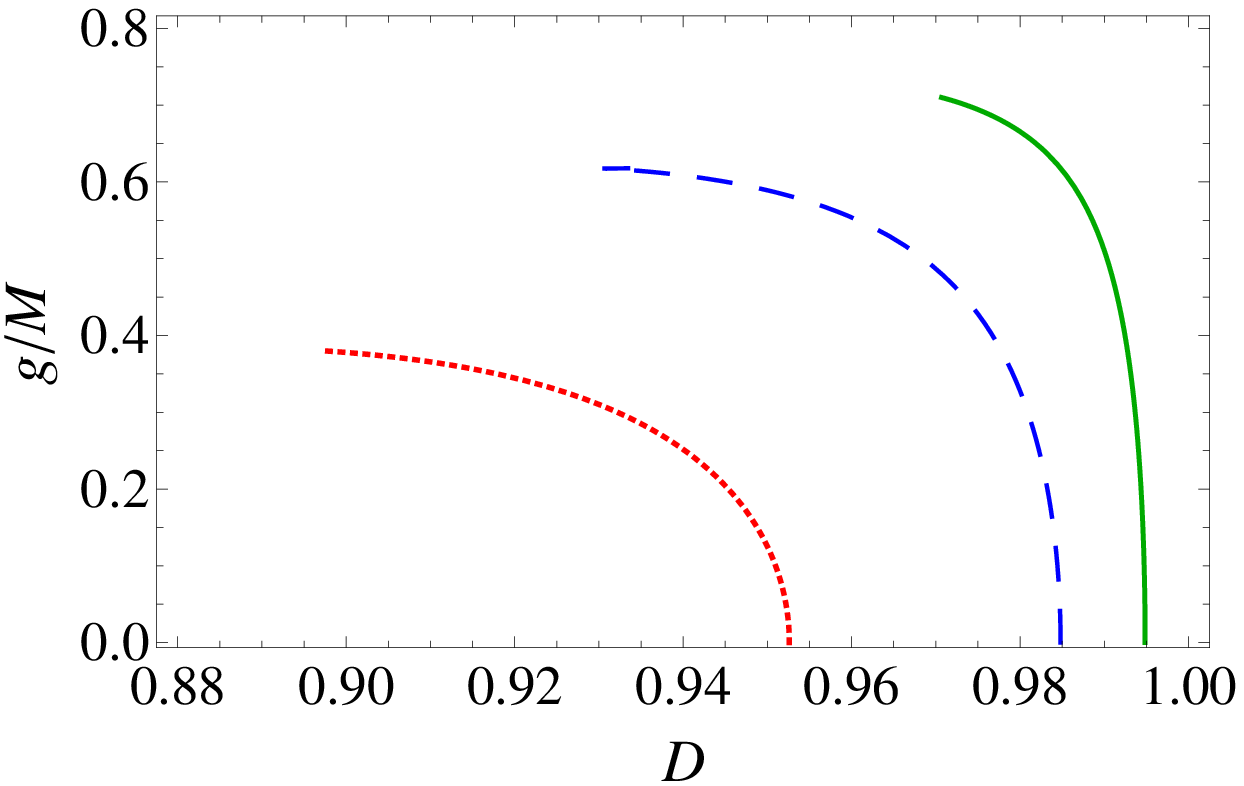} 
	\end{tabular}
	\caption{The magnetic charge parameter $g$ vs. observables $A$, $C$, and $D$ for the Bardeen black hole, for a nonrotating Bardeen black hole $a/M=0.0$ (solid black curve), for a rotating Bardeen black hole with $a/M=0.3$ (solid green curve), for $a/M=0.5$ (dashed blue curve), and for $a/M=0.8$ (dotted red curve).} \label{bardeen}
\end{figure*}

\begin{figure*}
	\begin{tabular}{c c c }
		\includegraphics[scale=0.45]{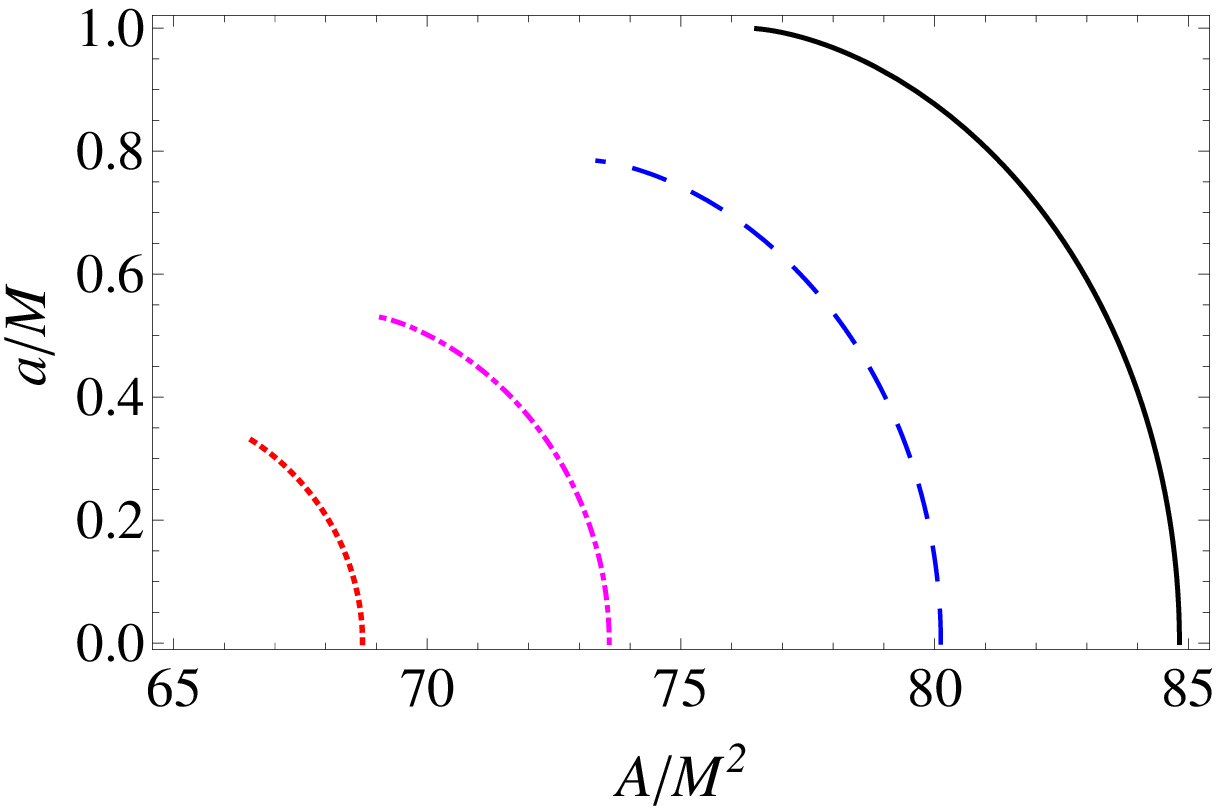}&
		\includegraphics[scale=0.45]{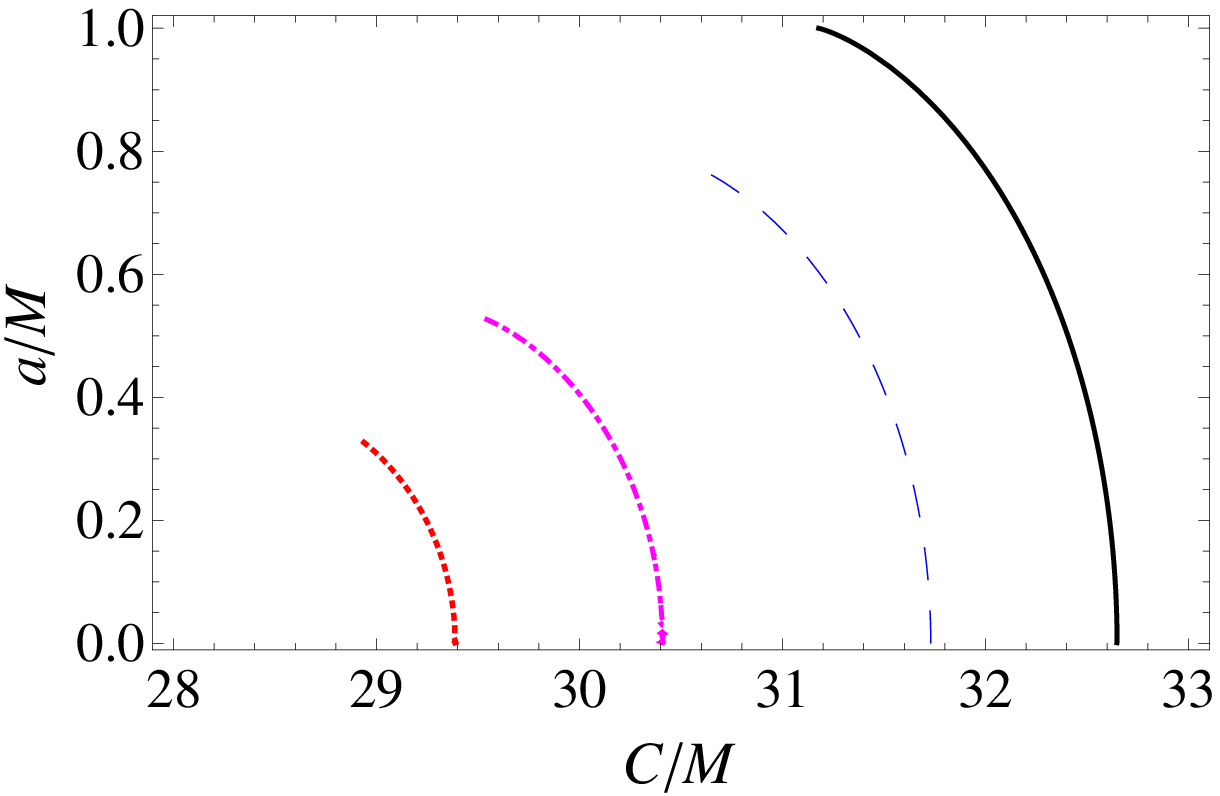}&
		\includegraphics[scale=0.45]{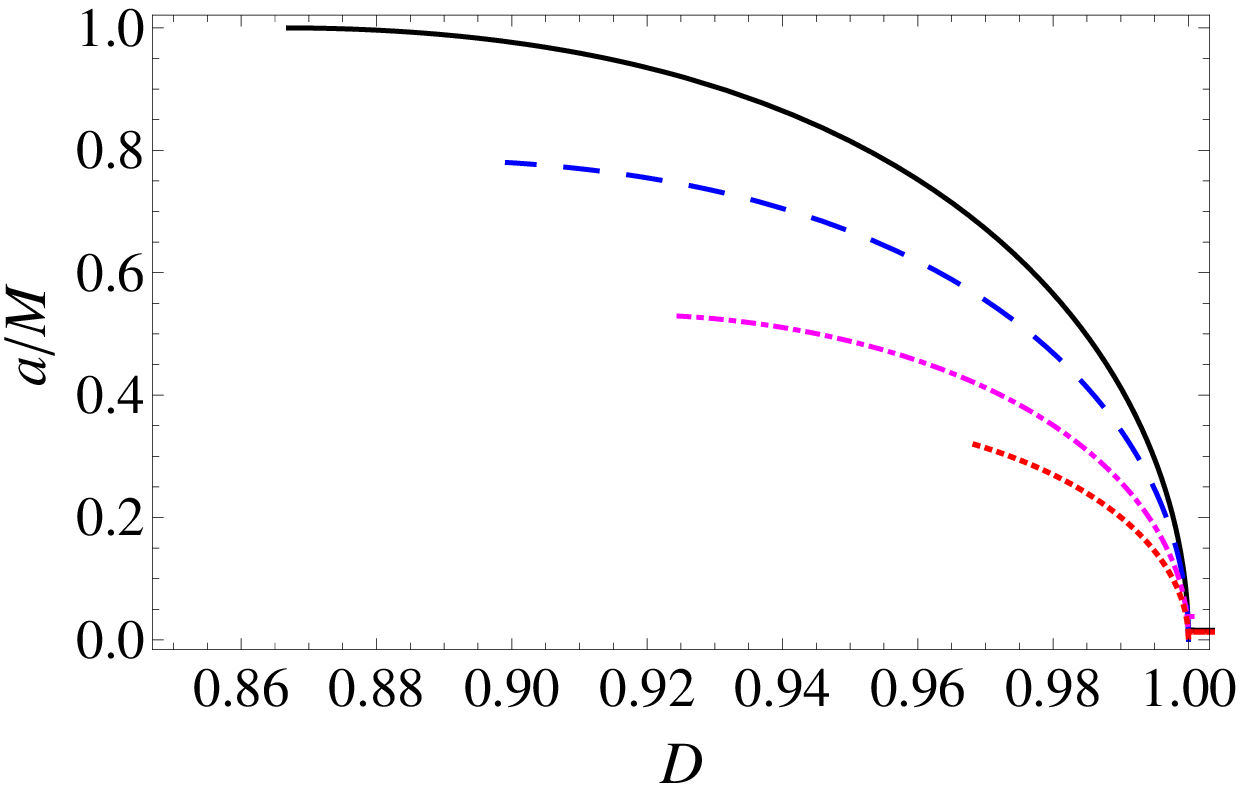} 
	\end{tabular}
	\caption{ The spin parameter $a$ vs. observables $A$, $C$, and $D$ for the Bardeen black hole, for $g/M=0.0$ (solid black curve), for $g/M=0.4$ (dashed blue curve), for $g/M=0.6$ (dotted dashed magenta curve), and for $g/M=0.7$ (dotted red curve).} \label{bardeen001}
\end{figure*}

\begin{figure}[h!]
	\includegraphics[scale=0.8]{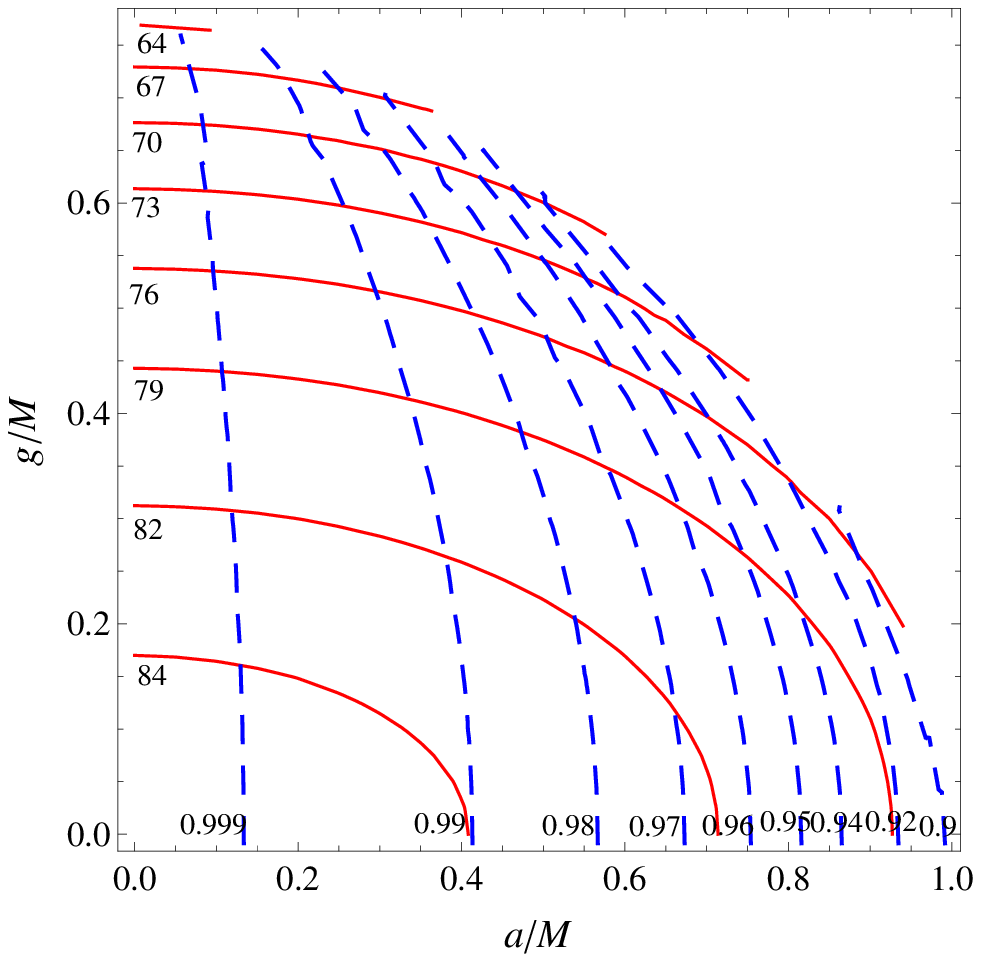} 
	\caption{Contour plot of the observables $A$ and $D$ in the plane $(a, g)$ for a Bardeen black hole. Each curve is labeled with the corresponding values of $A$ (solid red curve) and $D$ (dashed blue curve). } \label{bardeen01}
\end{figure}

The Bardeen black hole is an exact solution of the Einstein field equations coupled with nonlinear electrodynamics associated with the magnetic monopole charge  $g$ \citep{AyonBeato:1999rg}. The Kerr black hole can be recovered in the absence of the nonlinear electrodynamics ($g=0$). For the existence of a black hole, the allowed values of $a$ and $g$ are constrained and shown in Figure~\ref{BardeenNoBH}, and the extremal values of parameters correspond to those lying on the boundary line. The shadows of rotating Bardeen black holes get more distorted and  their sizes decrease due to the magnetic charge $g$ \citep{Abdujabbarov:2016hnw}.

The Bardeen black hole parameters $g$ and $a$ versus the observables $A$, $C$, and $D$ are depicted in Figures~\ref{bardeen} and \ref{bardeen001}, respectively. Within the allowed parameter space, they have a similar behavior to that of the Kerr$-$Newman black hole. The parameters decrease with increasing observables, however, for a near extremal black hole, parameters decrease comparatively slowly with increasing $D$. Further, the observables of a rotating Bardeen black hole are smaller when compared with the Kerr black hole for a given $a$, i.e., $A(g\neq 0)< A(g=0)$ and $D(g\neq 0)< D(g=0)$ (see Figure \ref{bardeen001}).  An interesting comparison between shadows of Bardeen and Kerr black holes shows that for some values of parameters, a Bardeen black hole ($M=1, a/M=0.5286, g/M=0.6$) casts a similar shadow to that of Kerr black hole ($M=0.9311, a/M=0.9189$) \citep{Tsukamoto:2014tja}. In this case, the observables for a Bardeen black hole are $A=69.1445, C=29.5269$, and $D=0.925402$, whereas for a Kerr black hole, they are $A=68.68015$, $C=29.4213$, and $D=0.925402$. Thus, the $A$ and $C$ for the two black holes differ by $0.671\%$ and $0.357\%$, respectively. The differences in their major and minor angular diameters are $0.0331\%$ and $2.158\%$, respectively. Figure \ref{bardeen01} shows the contour map of observables $A$ and $D$ for the rotating Bardeen black hole as a function of ($a, g$). 
In Table \ref{BH parameter}, we have shown the estimated values of Bardeen parameters $a$ and $g$ for given shadow observables $A$ and $D$, and compare them with the estimated values of other black hole parameters. Thus, from Figure~\ref{bardeen01} and Table \ref{BH parameter} it is clear that if $A$ and $D$ are known from the observations, this uniquely determines the $a$ and $g$.

\subsection{Rotating Nonsingular Black Hole}
The Bardeen regular black holes have a de-Sitter region at the core. Next, we consider a class of rotating regular black holes with asymptotically Minkowski cores \citep{simpson2020regular}, which have an additional parameter $k=q^2/{2M}>0$ due to nonlinear electrodynamics that deviates from Kerr and asymptotically ($r>>k$) goes over to a Kerr$-$Newman black hole \citep{Ghosh:2014pba}. Whilst this rotating regular black hole shares many properties with Bardeen rotating regular black holes, there is also a significant contrast, and for definiteness, we name it a rotating nonsingular black hole. It also belongs to the non-Kerr family with mass function 
\begin{equation}
m(r)=Me^{-k/r}.\label{NSmass}
\end{equation}
\begin{figure}[t!]
	\includegraphics[scale=0.7]{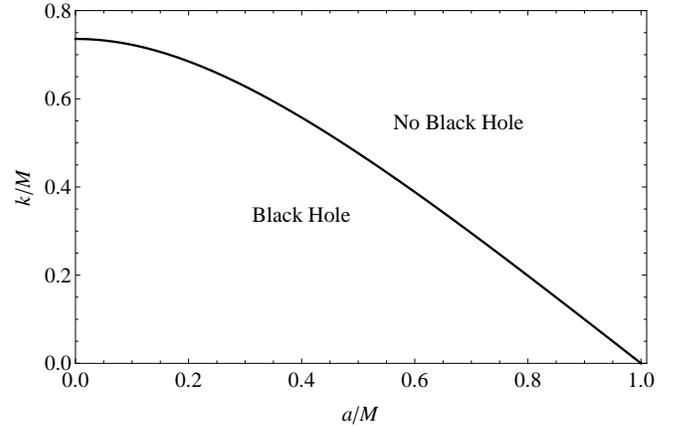}
	\caption{The allowed parametric space of $a$ and $k$ for the existence of rotating nonsingular black hole. The solid line corresponds to the extremal black hole with degenerate horizons.}\label{NonsingularNoBH}
\end{figure}
\begin{figure*}
	\begin{tabular}{c c c}
		\includegraphics[scale=0.45]{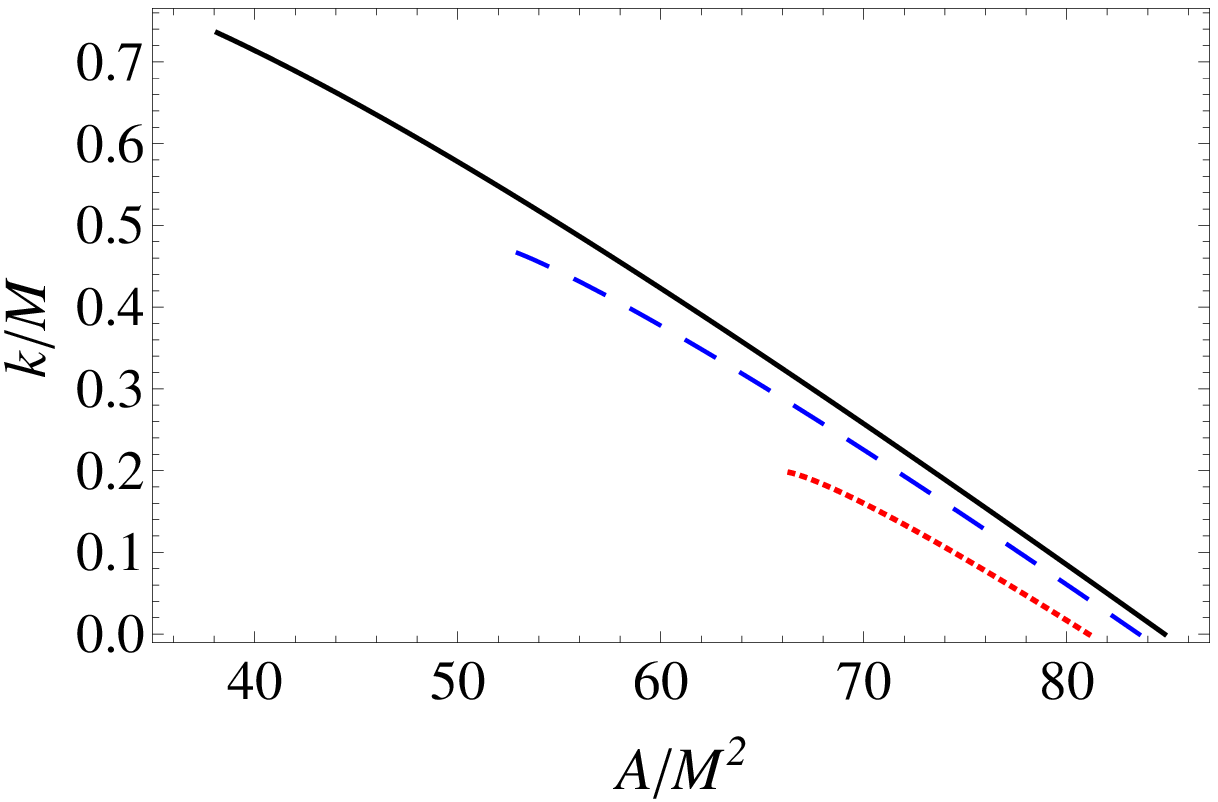}&
		\includegraphics[scale=0.45]{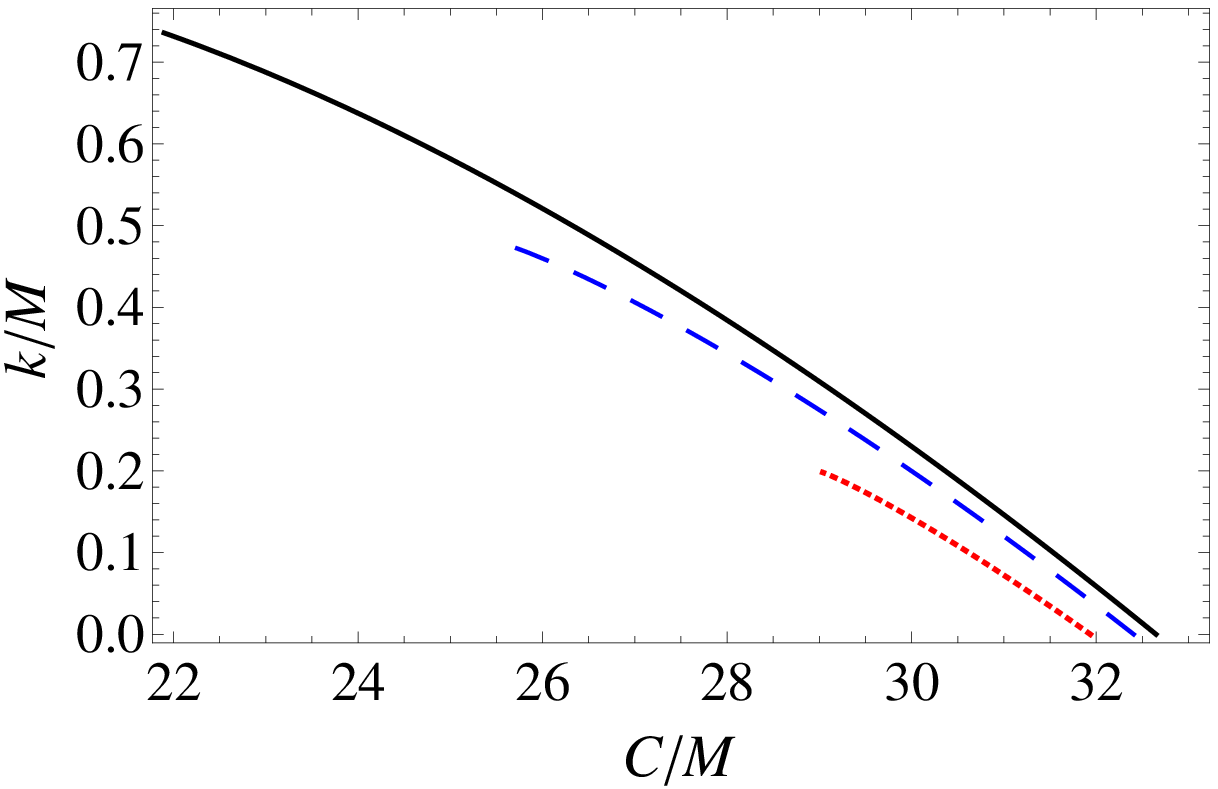}&
		\includegraphics[scale=0.45]{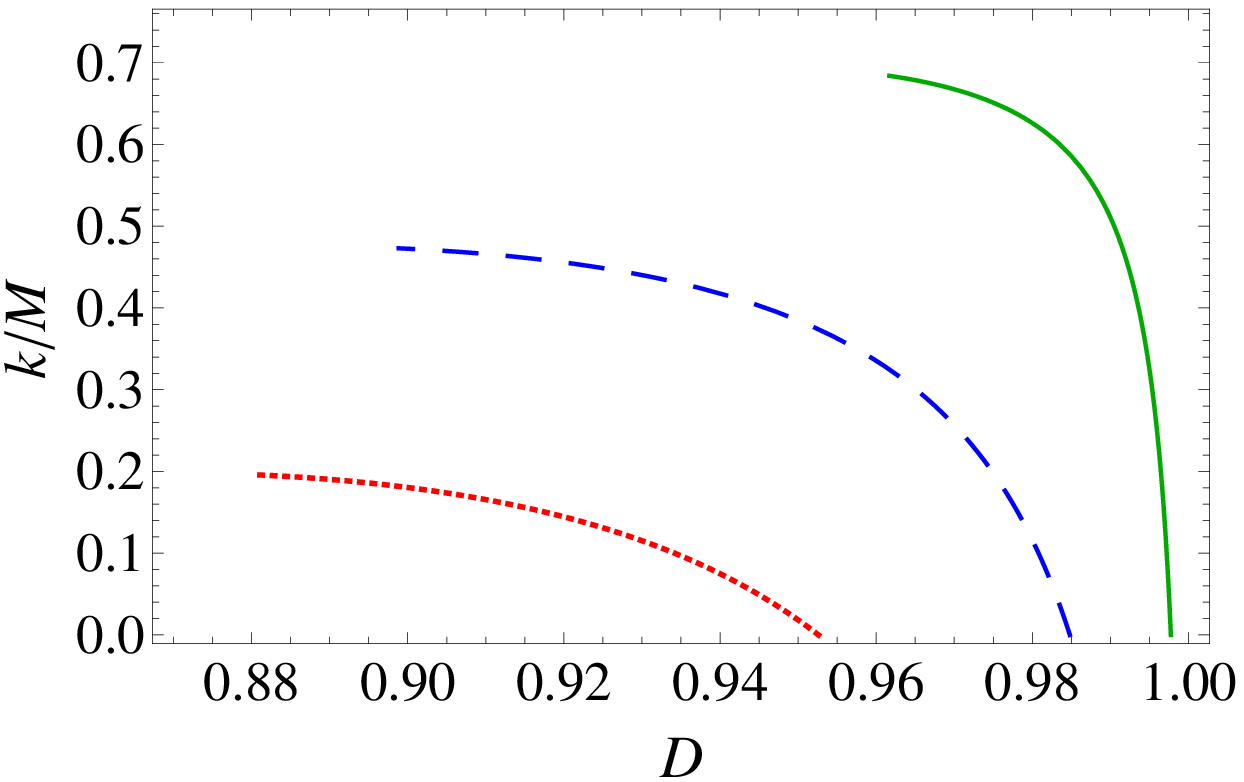}
	\end{tabular}
	\caption{Charge parameter $k$ vs. observables $A$, $C$, and $D$ for the nonsingular black hole, for $a/M=0.0$ (solid black curve), for $a/M=0.2$ (solid green curve), for $a/M=0.5$ (dashed blue curve) and for $a/M=0.8$ (dotted red curve).}\label{nonsingular}
\end{figure*}

\begin{figure*}
\begin{tabular}{c c c}
	\includegraphics[scale=0.45]{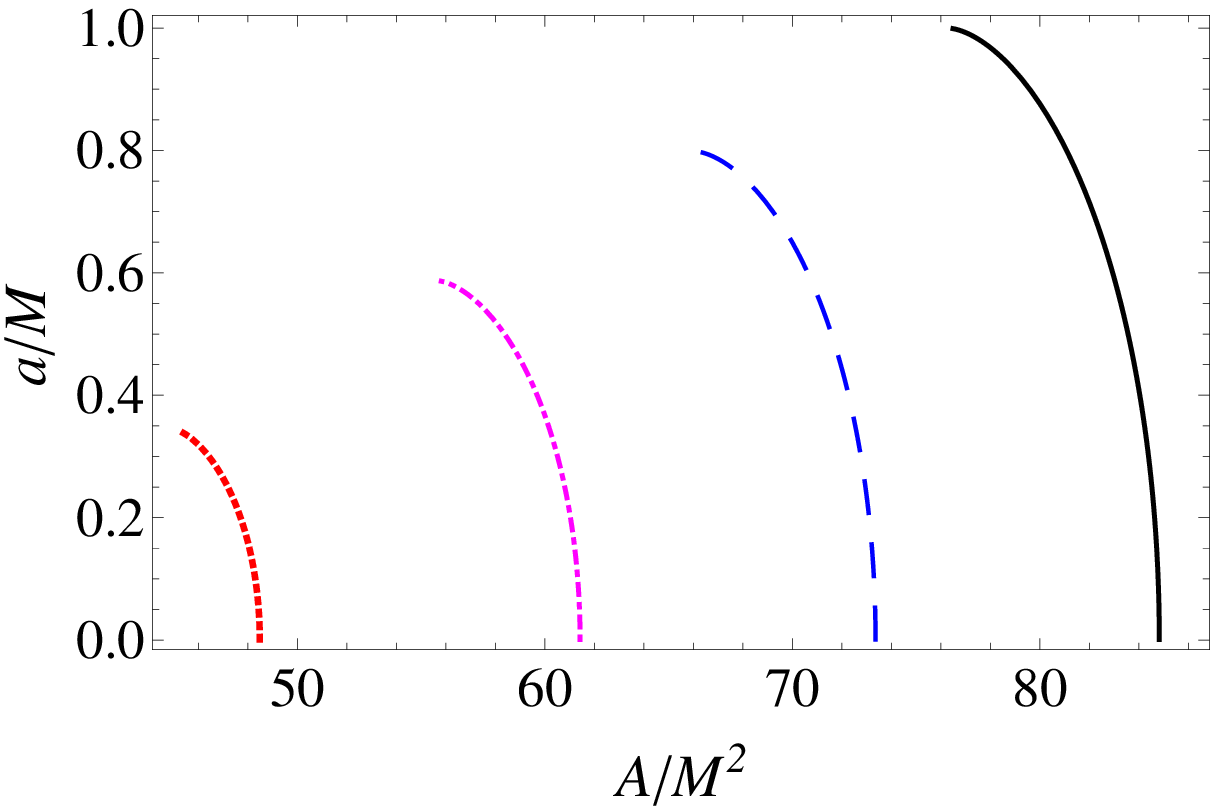}&
	\includegraphics[scale=0.45]{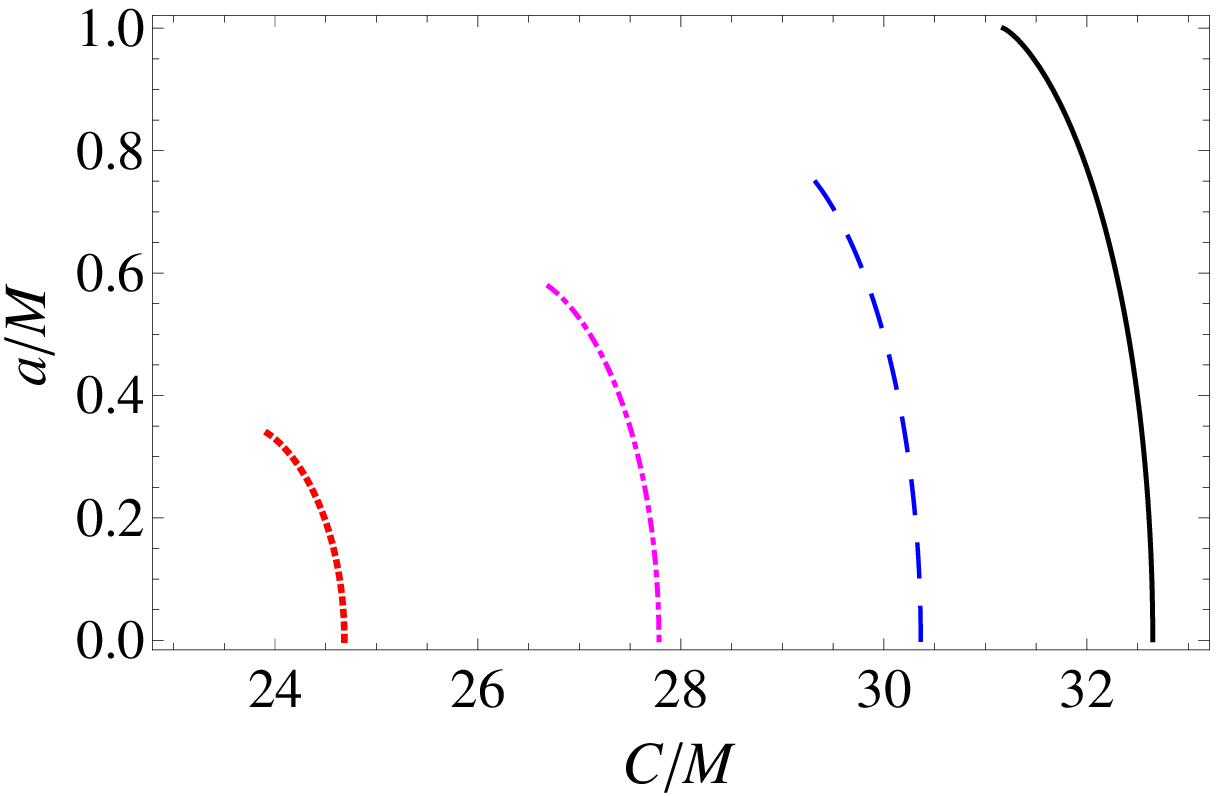}&
	\includegraphics[scale=0.45]{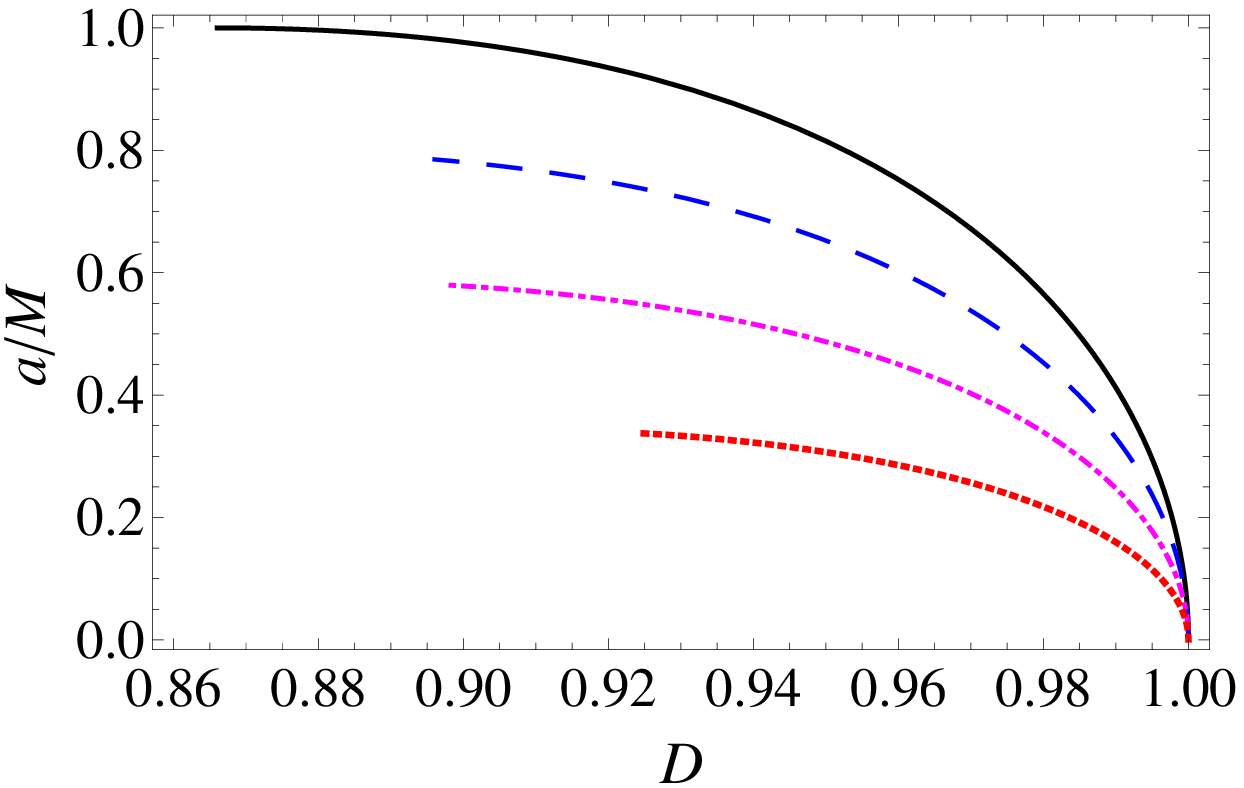} 
\end{tabular}
	\caption{Spin parameter $a$ vs. observables $A$, $C$, and $D$ for the nonsingular black hole, for $k/M=0.0$ (solid black curve), for $k/M=0.2$ (dashed blue curve), for $k/M=0.4$ (dotted dashed magenta curve), and for $k/M=0.6$ (dotted red curve).}\label{nonsingular001}
\end{figure*}

\begin{figure}[h!]
	\includegraphics[scale=0.8]{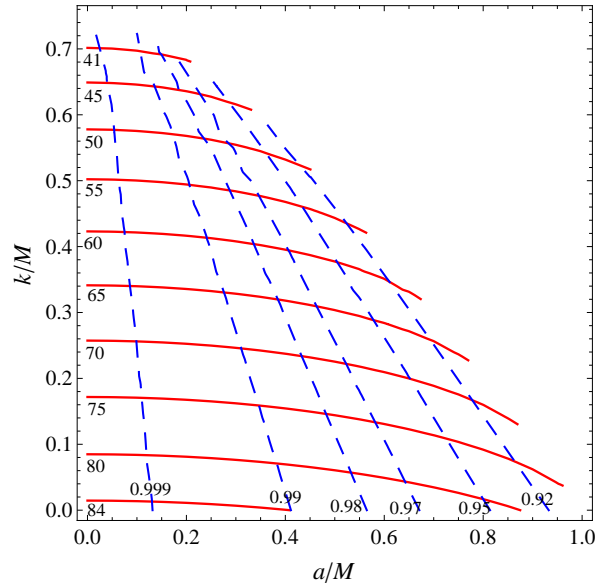} 
	\caption{Contours of constant $A$ and $D$ as a function of $(a, k)$ for a rotating nonsingular black hole. Each curve is labeled with the corresponding value of $A$ (solid red curve) and $D$ (dashed blue curve).}
	\label{nonsingular01}
\end{figure}

Figure \ref{NonsingularNoBH} shows the allowed values of parameters $a$ and $k$ for the black hole's existence. The effect of varying observables $A$, $C$, and $D$ on the inferred rotating nonsingular black hole parameters $k$ and $a$ are depicted in Figures~\ref{nonsingular} and \ref{nonsingular001}, respectively. The characteristic behavior is again similar to that for the Kerr$-$Newman, but the effect on $k$ is visible for both nonrotating and rotating nonsingular black holes (see Figures~\ref{nonsingular} and \ref{nonsingular001}). The estimated values of $k$ show a similar sharp decreasing behavior with increasing $A$ and $C$, whereas it slowly decreases with $D$ for near extremal black holes. The observables for rotating nonsingular black holes are examined in contrast with those for Kerr black holes in Figure~\ref{nonsingular001}, and for a fixed value of $a$ they turn out to be smaller. This indicates that shadows of rotating nonsingular black holes are smaller and more distorted than those of Kerr black holes \citep{Amir:2016cen}. Contour maps of $A$ and $D$ as a function of ($a,k$) are shown in Figure~\ref{nonsingular01}. We can easily determine the specific points where curves of constant $A$ and $D$ intersect each other in the black hole parameter space, yielding the unique values of $a$ and $k$. 
 
\subsection{Comparison of Estimated Black Hole Parameters}
Applying the method described in section 3, the numerical values of the three considered rotating black holes parameters, for a given shadow area $A$ and oblateness $D$, are summarized in the Table \ref{BH parameter}. Here, we compare the estimated black hole parameters for the three black holes. For a given shadow area $A$, we find that the spin parameter decreases with increasing oblateness $D$, and that for a fixed area $A$ and oblateness $D$ we obtain that the spin parameters are $a_{\text{NS}}>a_{\text{KN}}>a_{\text{Bardeen}}$ and the charge parameters are $Q>g>k$. For a fixed oblateness $D$, the charge parameters $Q$, $g$, and $k$ increase and the spin parameter $a$ decreases with a decrease in the area $A$. For small area $A$ and oblateness $D$, e.g., $A=55M^2$ and $D=0.92$, one could estimate parameters associated with only the rotating nonsingular black hole (see Table \ref{BH parameter}).

\section{Conclusion}\label{sect5}
The EHT has obtained the first image of the M87* black hole, and thus its shadow becomes an important probe of spacetime structure, parameter estimation, and testing gravity in the extreme region near the event horizon. Even though most of the available tests are consistent with general relativity, deviations from the Kerr black hole (or non-Kerr black hole) arising from modified theories of gravity are not ruled out  \citep{Johannsen:2011dh, Berti:2015itd}. These non-Kerr black holes, in Boyer$-$Lindquist coordinates, are defined by the metric (\ref{rotmetric}) with mass function $m(r)$, and Kerr black holes are included as special case when $m(r)=M$. 
In this paper, we have proposed observables, namely, shadow area ($A$), its circumference ($C$), and oblateness ($D$). The observables $A$ and $C$ characterize the size of the shadow, and $D$ defines its shape asymmetry. These observables are calculated for Sgr A* and M87*, assuming their Kerr nature, and we find that their angular diameters are approximately $52\, \mu$as and $39\,\mu$as, respectively, and decrease for a rapidly rotating black hole. This is consistent with other predicted results \citep{Falcke:2013ola,fish2014imaging,Brinkerink:2018bhw,Akiyama:2019cqa,Akiyama:2019bqs,Akiyama:2019eap}.\\
We highlight several other results that are obtained by our analysis. \\
\begin{enumerate}
	\item The method can estimate, at most, two parameters by using either $A$ or $C$ along with $D$; for example, the Kerr black hole parameters $a$ and $\theta_O$ can be estimated. In order to estimate a single parameter, we require any one of these observables.
	\item For given shadow observables, we have estimated parameters associated with Kerr$-$Newman ($a, Q$), rotating Bardeen ($a,g$), and rotating nonsingular ($a,k$) black holes. Here, our analysis assumes that the observer is in the equatorial plane, i.e., at a fixed inclination angle $\theta_O=\pi/2$.
	\item Our results for the considered black holes are consistent with existing results \citep{Tsukamoto:2014tja}.
	\item We have interpolated the numerical values of observables from integrals in Equations~(\ref{Area}) and (\ref{Circumference}) and used Equation (\ref{Oblateness}) to approximate these observables as polynomials in terms of the black hole parameters.
	\item Our analysis is applicable to a large variety of shadow shapes and does not require approximating the shadow as a circle.
\end{enumerate}
Thus, by comparing the theoretically calculated values of these observables with those obtained from the astrophysical observations, it is expected that one can completely determine information about a black hole. Our analysis is clearly different from other approaches but leads to the correct estimation of black hole parameters. Our framework can be extended to other classes of black holes.
 
A set of shadow observables can correspond to various black holes with different values of parameters involved (see Table~\ref{BH parameter}). Indeed, we find that a strong correlation between the spin and the deviation parameters from the Kerr solution makes it difficult to discern two black hole models with given shadow observables. It will be interesting to find new observables characterizing the shadows in the presence of an accretion disk; this and related projects are being investigated. 

\section{Acknowledgments}
S.G.G. would like to thank the DST INDO-SA bilateral project DST/INT/South Africa/P-06/2016 and also IUCAA, Pune for the hospitality while this work was being done. R.K. would like to thank UGC for providing SRF, and also Md Sabir Ali and Balendra Pratap Singh for fruitful discussions. The authors would like to thank the anonymous reviewer for providing insightful comments which immensely helped to improve the paper. 

\appendix
\section{Analytic form of Observables}\label{Appendix1}
The celestial coordinates $\alpha$ and $\beta$ can be calculated via Equation~(\ref{impactparameter}) for a given mass function, and in turn, they help us to calculate observables $A$, $C$, and $D$ numerically. Here, we present an approximate and analytic form of $A$, $C$, and $D$ obtained from the best fit of the numerical data for the three discussed rotating black holes. For a Kerr$-$Newman black hole it yields

\begin{eqnarray}
\frac{A(a,Q)}{M^2} &=&84.823 - 0.0241486 \frac{a}{M} - 3.92067\frac{a^2}{M^2} - 7.64929\frac{a^3}{M^3} + 31.6438\frac{a^4}{M^4} -70.2995\frac{a^5}{M^5}\nonumber\\
& +&73.3373 \frac{a^6}{M^6} - 30.7606\frac{a^7}{M^7} - 2.21765\frac{ Q}{M} - 6.0038\frac{ a Q}{M^2} + 52.4736\frac{ a^2 Q}{M^3} - 
210.27\frac{ a^3 Q}{M^4} \nonumber\\
& +& 415.819 \frac{a^4 Q}{M^5} - 509.279 \frac{a^5 Q}{M^6} + 356.297 \frac{a^6 Q}{M^7} + 11.9166 \frac{Q^2}{M^2} + 43.6266 \frac{a Q^2}{M^3} -342.836 \frac{a^2 Q^2}{M^4} \nonumber\\
&+& 1144.53 \frac{a^3 Q^2}{M^5} - 1460.97 \frac{a^4 Q^2}{M^6} + 988.7 \frac{a^5 Q^2}{M^7} - 287.055 \frac{Q^3}{M^3} - 110.416 \frac{a Q^3}{M^4} + 786.043 \frac{a^2 Q^3}{M^5} \nonumber\\
&-&2403.56 \frac{a^3 Q^3}{M^6} + 1916.18\frac{ a^4 Q^3}{M^7}  + 1064.15 \frac{Q^4}{M^4} + 116.604 \frac{a Q^4}{M^5} - 757.586 \frac{a^2 Q^4}{M^6} + 2203.57 \frac{a^3 Q^4}{M^7}\nonumber\\
&-& 2257.51 \frac{Q^5}{M^5} - 44.1061 \frac{a Q^5}{M^6} + 253.501 \frac{a^2 Q^5}{M^7 } + 2735.71 \frac{Q^6}{M^6} - 1769.27 \frac{Q^7}{M^7},\nonumber\\
\frac{C(a,Q)}{M}&=&32.6484 - 0.00333693 \frac{a}{M} - 0.801683 \frac{a^2}{M^2} - 1.00595 \frac{a^3}{M^3} + 4.08401 \frac{a^4}{M^4} - 
9.22737 \frac{a^5}{M^5} \nonumber\\
& +& 9.63857 \frac{a^6}{M^6} - 4.08849 \frac{a^7}{M^7} - 0.252252 \frac{Q}{M} - 0.952349 \frac{a Q}{M^2} + 7.90764 \frac{a^2 Q}{M^3} - 30.7126 \frac{a^3 Q}{M^4} \nonumber\\
&+&55.564 \frac{a^4 Q}{M^5} -65.4834 \frac{a^5 Q}{M^6} + 44.8428 \frac{a^6 Q}{M^7} - 0.986894 \frac{Q^2}{M^2} + 7.56037 \frac{a Q^2}{M^3} -56.8018\frac{ a^2 Q^2}{M^4}\nonumber\\
& +& 187.173 \frac{a^3 Q^2}{M^5} - 212.661 \frac{a^4 Q^2}{M^6} + 137.518 \frac{a^5 Q^2}{M^7}  - 30.7211 \frac{Q^3}{M^3} - 20.9274 \frac{a Q^3}{M^4} + 141.551 \frac{a^2 Q^3}{M^5}\nonumber\\
&-&438.524 \frac{a^3 Q^3}{M^6} + 299.757 \frac{a^4 Q^3}{M^7}  + 108.328 \frac{Q^4}{M^4} + 24.2099 \frac{a Q^4}{M^5} - 149.282 \frac{a^2 Q^4}{M^6} + 450.497 \frac{a^3 Q^4}{M^7}\nonumber\\
&-& 219.805 \frac{Q^5}{M^5} - 10.0047 \frac{a Q^5}{M^6} + 54.27 \frac{a^2 Q^5}{M^7}+ 250.578 \frac{Q^6}{M^6} - 151.118 \frac{Q^7}{M^7},\nonumber\\
D(a,Q)&=& 1. - 0.000544168 \frac{a}{M} - 0.0377214 \frac{a^2}{M^2} - 0.172164 \frac{a^3}{M^3} + 0.719728 \frac{a^4}{M^4} - 
1.57686 \frac{a^5}{M^5} \nonumber\\
&+& 1.6425 \frac{a^6}{M^6}- 0.683473 \frac{a^7}{M^7} - 0.160047 \frac{a Q}{M^2} + 1.38395 \frac{a^2 Q}{M^3} - 5.40696 \frac{a^3 Q}{M^4} +10.432 \frac{a^4 Q}{M^5} \nonumber\\
&-&  12.6347 \frac{a^5 Q}{M^6} + 8.78616 \frac{a^6 Q}{M^7}  + 1.19921 \frac{a Q^2}{M^3} - 9.34042 \frac{a^2 Q^2}{M^4} + 30.3103 \frac{a^3 Q^2}{M^5} \nonumber\\
&-& 37.3007 \frac{a^4 Q^2}{M^6} + 24.7655 \frac{a^5 Q^2}{M^7} - 3.14047 \frac{a Q^3}{M^4} + 22.4071 \frac{a^2 Q^3}{M^5} -65.711 \frac{a^3 Q^3}{M^6} + 49.9938 \frac{a^4 Q^3}{M^7}\nonumber\\
&+& 3.4426 \frac{a Q^4}{M^5} - 22.7217 \frac{a^2 Q^4}{M^6} + 62.2473 \frac{a^3 Q^4}{M^7}- 1.3531 \frac{a Q^5}{M^6} + 8.14017 \frac{a^2 Q^5}{M^7}.\label{series}
\end{eqnarray}

Clearly, $A$, $C$, and $D$ are functions of spin $a$ and charge $Q$. For the Bardeen black hole, they depend upon the magnetic charge $g$ in addition to $a$, and are given by

\begin{eqnarray}
\frac{A(a,g)}{M^2}&=& 84.823 - 0.024064 \frac{a}{M} - 3.92154 \frac{a^2}{M^2} - 7.64519 \frac{a^3}{M^3} + 31.6335 \frac{a^4}{M^4} - 70.2853 \frac{a^5}{M^5}  \nonumber\\
&+& 73.327 \frac{a^6}{M^6} -30.7577 \frac{a^7}{M^7}+0.0715772 \frac{g}{M}+ 10.1889 \frac{ag}{M^2} - 88.754 \frac{a^2g}{M^3} + 310.385 \frac{a^3g}{M^4} \nonumber\\
&-& 492.683 \frac{a^4g}{M^5} + 323.536\frac{a^5g}{M^6}+9.27673\frac{a^6 g}{M^7}- 33.5425 \frac{g^2}{M^2} - 157.492 \frac{ag^2}{M^3} + 1350.8 \frac{a^2g^2}{M^4}\nonumber\\
&-&4604.46 \frac{a^3g^2}{M^5}+ 7243.43 \frac{a^4g^2}{M^6}-5690.2 \frac{a^5 g^2}{M^7}+ 77.0126\frac{g^3}{M^3}+ 837.866 \frac{ag^3}{M^4}-7045.26 \frac{a^2g^3}{M^5}\nonumber\\
&+& 21471.2 \frac{a^3g^3}{M^6} -28065.2 \frac{a^4 g^3}{M^7} - 483.12 \frac{g^4}{M^4} -1957.29 \frac{ag^4}{M^5}+16337.0 \frac{a^2 g^4}{M^6} - 42607.5 \frac{a^3 g^4}{M^7}\nonumber\\
&+&1519.55 \frac{g^5}{M^5} + 1913.41 \frac{a g^5}{M^6}-17719.5 \frac{a^2 g^5}{M^7}-2628.46\frac{g^6}{M^6} +289.67 \frac{ag^6}{M^7}- 2343.44 \frac{g^7}{M^7},\nonumber\\
\frac{C(a,g)}{M}&=& 32.6484 - 0.00340862 \frac{a}{M} - 0.801464 \frac{a^2}{M^2} - 1.00523 \frac{a^3}{M^3} + 4.07939 \frac{a^4}{M^4} - 9.21854 \frac{a^5}{M^5} \nonumber\\
&+&  9.63115 \frac{a^6}{M^6}-4.08615 \frac{a^7}{M^7}+0.08745 \frac{g}{M} + 1.41767 \frac{ag}{M^2} - 12.2238 \frac{a^2g}{M^3} + 41.4626 \frac{a^3g}{M^4} \nonumber\\
&-& 62.7011 \frac{a^4 g}{M^5}+ 38.1636 \frac{a^5g}{M^6} +4.13287 \frac{a^6 g}{M^7}-8.15235 \frac{g^2}{M^2}-23.2831 \frac{ag^2}{M^3} + 198.21 \frac{a^2g^2}{M^4} \nonumber\\
&-& 660.393 \frac{a^3g^2}{M^5} + 1000.05 \frac{a^4g^2}{M^6}-765.47 \frac{a^5 g^2}{M^7}+30.7989 \frac{g^3}{M^3} + 128.559 \frac{ag^3}{M^4}- 1084.82 \frac{a^2g^3}{M^5}  \nonumber\\
&+&3228.88 \frac{a^3g^3}{M^6} -4041.99 \frac{a^4 g^3}{M^7}-173.259 \frac{g^4}{M^4} - 308.254 \frac{ag^4}{M^5} +2625.19 \frac{a^2 g^4}{M^6} - 6718.79 \frac{a^3 g^4}{M^7} \nonumber\\
&+& 521.402 \frac{g^5}{M^5}+ 302.965\frac{ag^5}{M^6}-883.676 \frac{g^6}{M^6}-28.9381 \frac{a g^6}{M^7}+782.985 \frac{g^7}{M^7},\nonumber\\
D(a,g)&=& 1. - 0.000544168 \frac{a}{M} - 0.0377214 \frac{a^2}{M^2} - 0.172164 \frac{a^3}{M^3} + 0.719728 \frac{a^4}{M^4} -  1.57686 \frac{a^5}{M^5}  \nonumber\\
&+& 1.6425 \frac{a^6}{M^6}-0.683473  \frac{a^7}{M^7} + 0.28401 \frac{ag}{M^2} - 2.47242 \frac{a^2g}{M^3} + 8.78248 \frac{a^3g}{M^4} -  14.4606 \frac{a^4g}{M^5} \nonumber\\
&+&  10.5194 \frac{a^5g}{M^6}-1.18703 \frac{a^6g}{M^7} - 4.30871 \frac{ag^2}{M^3} + 36.6622 \frac{a^2g^2}{M^4} - 124.886 \frac{a^3g^2}{M^5} +  197.753 \frac{a^4g^2}{M^6}\nonumber\\
&-&155.984 \frac{a^5 g^2}{M^7}+22.8559 \frac{ag^3}{M^4} - 188.187 \frac{a^2g^3}{M^5} + 573.07 \frac{a^3g^3}{M^6}-751.811 \frac{a^4g^3}{M^7} - 54.0374 \frac{ag^4}{M^5} \nonumber\\
&+& 433.096 \frac{a^2g^4}{M^6}-1125.49 \frac{a^3 g^4}{M^7} + 55.5103 \frac{ag^5}{M^6}-466.675  \frac{a^2 g^5}{M^7} -13.6694 \frac{a g^6}{M^7}.\label{series2}
\end{eqnarray}

For the rotating nonsingular black hole, they are functions of $a$ and $k$ and read as
\begin{eqnarray}
\frac{A(a,k)}{M^2}&=& 84.823 -0.024298\frac{a}{M}-3.91914 \frac{a^2}{M^2} -7.6566  \frac{a^3}{M^3}+31.6622\frac{a^4}{M^4}-70.3248  \frac{a^5}{M^5} \nonumber\\
& +&73.3553 \frac{a^6}{M^6} -30.7658 \frac{a^7}{M^7}- 56.3846  \frac{k}{M}+0.53105  \frac{ak}{M^2}-20.4783 \frac{a^2k}{M^3}+139.145  \frac{a^3k}{M^4} \nonumber\\
&-& 410.288 \frac{a^4k}{M^5} +467.355\frac{a^5k}{M^6} -221.434 \frac{a^6k}{M^7}- 5.25777 \frac{k^2}{M^2} -2.43019 \frac{ak^2}{M^3} +40.2944\frac{a^2k^2}{M^4}\nonumber\\
&-&663.862 \frac{a^3k^2}{M^5} +1494.64 \frac{a^4k^2}{M^6} -735.119 \frac{a^5 k^2}{M^7}+ 5.63201 \frac{k^3}{M^3}- 1.5995\frac{ak^3}{M^4}  +37.8412 \frac{a^2k^3}{M^5}\nonumber\\
&+& 901.603 \frac{a^3k^3}{M^6} -1720.55\frac{a^4k^3}{M^7}- 14.1295 \frac{k^4}{M^4}+3.844\frac{ak^4}{M^5} -239.61 \frac{a^2k^4}{M^6},\nonumber\\
\frac{C(a,k)}{M}&=&32.6484 -0.00339904  \frac{a}{M}-0.800873 \frac{a^2}{M^2} -1.00981 \frac{a^3}{M^3}+4.09283 \frac{a^4}{M^4}-9.23782  \frac{a^5}{M^5} \nonumber\\
&+& 9.64471 \frac{a^6}{M^6} - 4.0899 \frac{a^7}{M^7}-10.8185\frac{k}{M} + 0.699341 \frac{ak}{M^2} -6.44615  \frac{a^2k}{M^3}  +22.6968  \frac{a^3k}{M^4}\nonumber\\
& -&65.7711 \frac{a^4k}{M^5} +70.7009\frac{a^5k}{M^6} -32.8498 \frac{a^6 k}{M^7} - 3.24875 \frac{k^2}{M^2} -7.91142 \frac{ak^2}{M^3} +38.7444 \frac{a^2k^2}{M^4}  \nonumber\\
&-& 118.877\frac{a^3k^2}{M^5} +268.241 \frac{a^4k^2}{M^6} -123.612\frac{a^5k^2}{M^7} + 1.99375 \frac{k^3}{M^3}  +27.4976 \frac{ak^3}{M^4}  -111.238 \frac{a^2k^3}{M^5}\nonumber\\
&+& 169.666 \frac{a^3 k^3}{M^6} - 334.206 \frac{a^4 k^3}{M^7}  - 5.67007 \frac{k^4}{M^4} -29.498 \frac{ak^4}{M^5} +82.1988 \frac{a^2k^4}{M^6} ,\nonumber\\
D(a,k)&=& 1. - 0.000544168 \frac{a}{M}  - 0.0377214 \frac{a^2}{M^2}  - 0.172164 \frac{a^3}{M^3}  + 0.719728 \frac{a^4}{M^4}  - 1.57686 \frac{a^5}{M^5} \nonumber\\
&+& 1.6425 \frac{a^6}{M^6} - 0.683473 \frac{a^7}{M^7}   +  0.014452 \frac{ak}{M^2}  - 0.490382 \frac{a^2k}{M^3}  + 3.7850 \frac{a^3k}{M^4}  - 10.804 \frac{a^4k}{M^5} \nonumber\\
&+& 11.709 \frac{a^5k}{M^6} - 5.37221 \frac{a^6k}{M^7}   - 
0.071081 \frac{ak^2}{M^3}  + 1.20427 \frac{a^2k^2}{M^4}  - 19.6311 \frac{a^3k^2}{M^5} +43.3697 \frac{a^4k^2}{M^6}  \nonumber\\
&-& 19.823 \frac{a^5k^2}{M^7}+ 0.0486 \frac{ak^3}{M^4}  + 1.1621 \frac{a^2k^3}{M^5}  + 28.0021 \frac{a^3k^3}{M^6}  - 
52.7191 \frac{a^4k^3}{M^7}  + 0.1211 \frac{ak^4}{M^5} \nonumber\\
& -& 7.30097 \frac{a^2k^4}{M^6}. \label{nonsingularfunc}
\end{eqnarray}
Here, we have presented the series up to $\mathcal{O}(M^{-7})$. The nonrotating black hole ($a=0$) casts a perfect circular shadow \citep{Synge:1966,Chandrasekhar:1985kt}, which is also fully consistent  with  Equations~(\ref{series})-(\ref{nonsingularfunc}), i.e., $D(0,Q)=D(0,g)=D(0,k)=1$. 

\section{Observables in Association With Noisy Data}
\begin{figure*}
	\begin{tabular}{c c c}
		\includegraphics[scale=0.55]{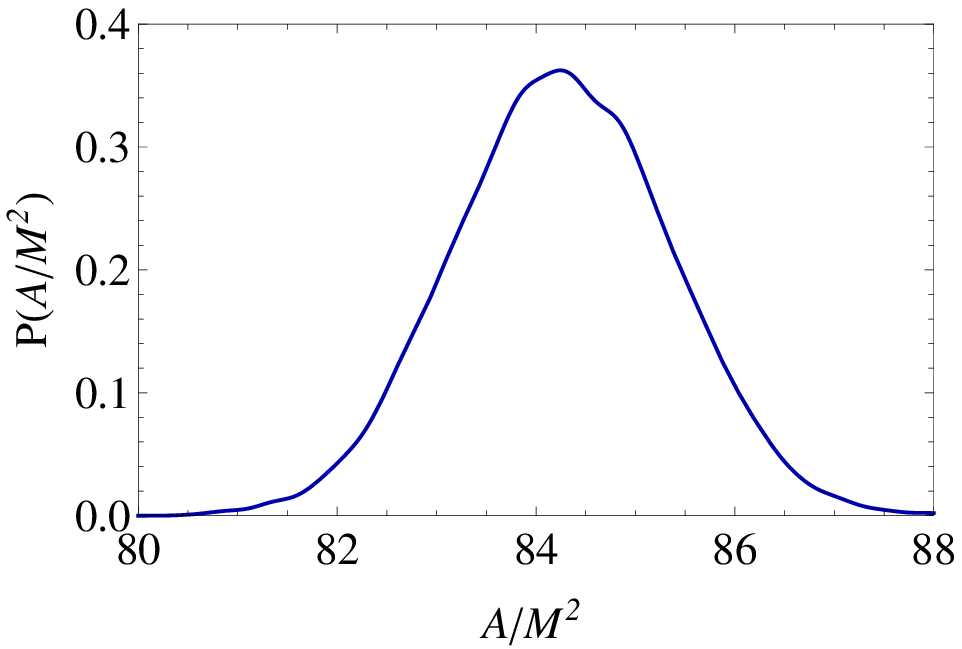}&
		\includegraphics[scale=0.55]{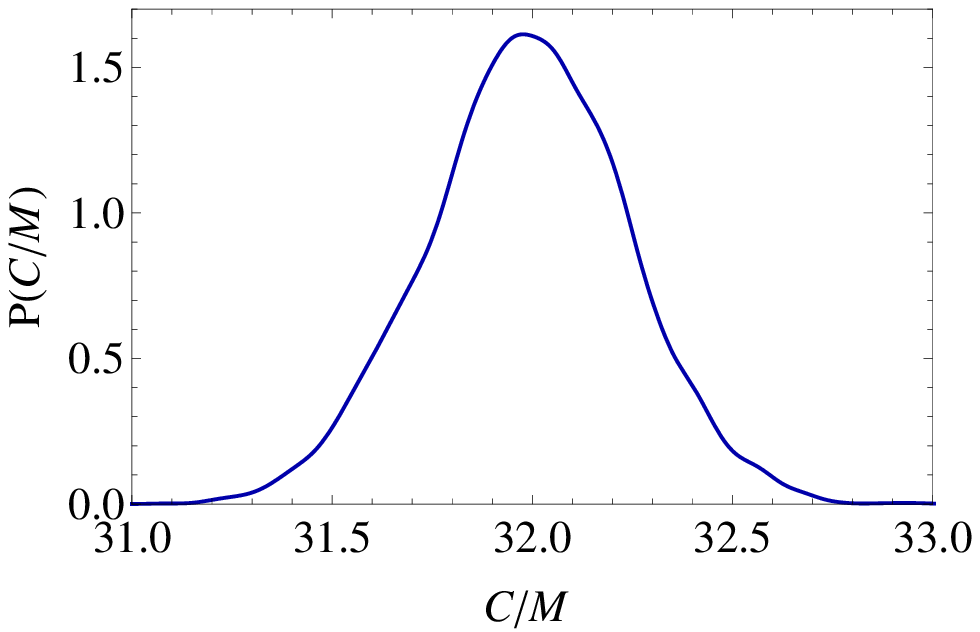}&
		\includegraphics[scale=0.55]{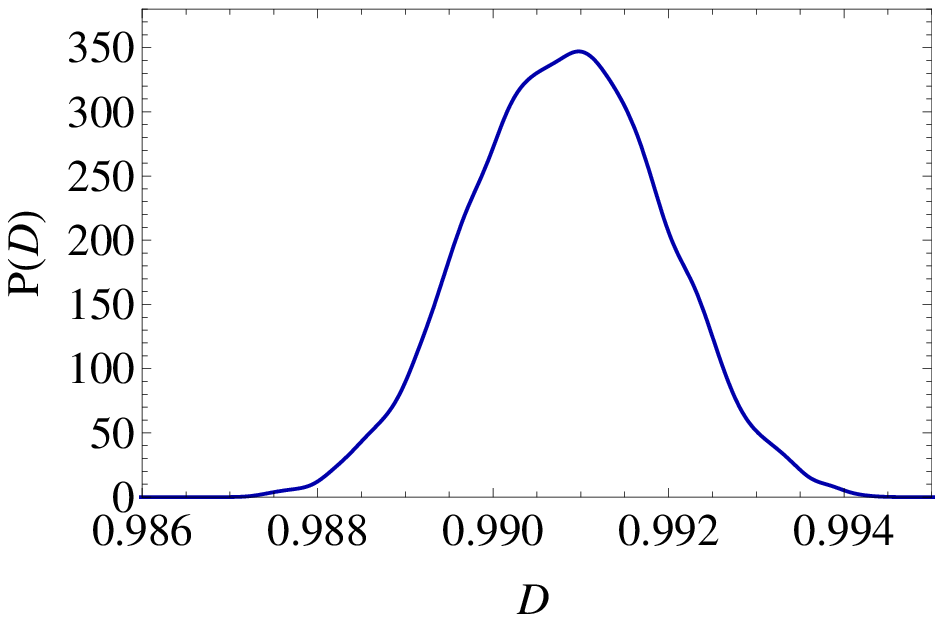} 
	\end{tabular}
	\caption{ Probability density distribution of shadow observables for perturbed Kerr black hole shadows.}\label{Noise}
\end{figure*}

The observables $A$, $C$, and $D$ are described in terms of the celestial coordinates $(\alpha,\beta)$, which are easy to calculate for a given black hole. Astronomical observations may not give a sharp shadow boundary demarcating the bright and dark regions; rather there will be intrinsic uncertainty in determining the shadow boundary because of noise in the observational data. In such observational data, we consider the set of visibility data points ($\alpha_i,\beta_i$) along the hazy shadow boundary. The geometric center $(\alpha_G,\beta_G)$ of the apparent shadow reads
\begin{equation}
\alpha_G=\frac{1}{N}\sum_{i=1}^{N}\alpha_i;\;\;\;\;\;\;\ \beta_G=\frac{1}{N}\sum_{i=1}^{N}\beta_i,
\end{equation}
where $N$ is the total number of data points. In the coordinate system centered at $(\alpha_G,\beta_G)$, the shadow boundary can be parameterized by $(\alpha'_i,\beta'_i)$
\begin{equation}
\alpha'_i=\alpha_i-\alpha_G;\;\;\;\;\;\;\ \beta'_i=\beta_i-\beta_G.
\end{equation}
Thus, we can calculate the shadow observables, $A$ and $C$, respectively, as \begin{equation}
A=\sum_{i=1}^N \frac{|\beta'_{i-1}+\beta'_i|}{2}|\alpha'_i-\alpha'_{i-1}|,
\end{equation}
and 
\begin{equation}
C=\sum_{i=1}^N \left((\alpha'_i-\alpha'_{i-1})^2+(\beta'_i-\beta'_{i-1})^2\right)^{1/2},
\end{equation}
where $\alpha'_0=0$ or $\alpha_0=\alpha_G$ and data points are arranged such that $|\alpha'_i|\geq |\alpha'_{i-1}|$. In this case, the oblateness $D$ becomes
\begin{equation}
D=\frac{\alpha'_r-\alpha'_l}{\beta'_t-\beta'_b},
\end{equation}
where $(\alpha'_l,0)$ and $(\alpha'_r,0)$ are, respectively, coordinates for the left and right edges of the shadow boundary, and $(\alpha'_t,\beta'_t)$ and $(\alpha'_b,\beta'_b)$ are for the top and bottom edges. Let's consider a contorted Kerr black hole shadow, whose boundary is artificially perturbed from a Kerr shadow ($\alpha_i, \beta_i$) and parameterized by ($\alpha_i+\epsilon_i, \beta_i+\epsilon_i$), where $\epsilon_i$ are random real numbers arbitrarily chosen in the interval [-0.01, 0.01] mimicking the noise in the observational data. For a Kerr black hole reference shadow ($a=0.3M$), we have developed a large number of synthetically perturbed shadows with the $10^4$ random noise distributions. The probability density functions $P$ for each observable are shown in Figure~\ref{Noise}, which are centered around the mean values of the corresponding shadow observable. It is seen that for a Kerr black hole shadow  ($a=0.3M$), we have $A=84.3889M^2$, $C=32.5649M$, and $D=0.994847$, whereas mean observables for the contorted Kerr black hole shadows are $A=84.2628M^2$, $C=31.9872M$, and $D=0.990859$. 

\bibliography{APJ}{}

\begin{thebibliography}{}
\expandafter\ifx\csname natexlab\endcsname\relax\def\natexlab#1{#1}\fi
\providecommand{\url}[1]{\href{#1}{#1}}
\providecommand{\dodoi}[1]{doi:~\href{http://doi.org/#1}{\nolinkurl{#1}}}
\providecommand{\doeprint}[1]{\href{http://ascl.net/#1}{\nolinkurl{http://ascl.net/#1}}}
\providecommand{\doarXiv}[1]{\href{https://arxiv.org/abs/#1}{\nolinkurl{https://arxiv.org/abs/#1}}}

\bibitem[{Abdujabbarov {et~al.}(2016)Abdujabbarov, Amir, Ahmedov, \&
  Ghosh}]{Abdujabbarov:2016hnw}
Abdujabbarov, A., Amir, M., Ahmedov, B., \& Ghosh, S.~G. 2016, Phys. Rev., D93,
  104004, \dodoi{10.1103/PhysRevD.93.104004}

\bibitem[{Abdujabbarov {et~al.}(2015{\natexlab{a}})Abdujabbarov, Atamurotov,
  Dadhich, Ahmedov, \& Stuchlik}]{Abdujabbarov:2015rqa}
Abdujabbarov, A., Atamurotov, F., Dadhich, N., Ahmedov, B., \& Stuchlik, Z.
  2015{\natexlab{a}}, Eur. Phys. J., C75, 399,
  \dodoi{10.1140/epjc/s10052-015-3604-5}

\bibitem[{Abdujabbarov {et~al.}(2015{\natexlab{b}})Abdujabbarov, Rezzolla, \&
  Ahmedov}]{Abdujabbarov:2015xqa}
Abdujabbarov, A.~A., Rezzolla, L., \& Ahmedov, B.~J. 2015{\natexlab{b}}, Mon.
  Not. Roy. Astron. Soc., 454, 2423, \dodoi{10.1093/mnras/stv2079}

\bibitem[{Akiyama {et~al.}(2019{\natexlab{a}})}]{Akiyama:2019cqa}
Akiyama, K., {et~al.} 2019{\natexlab{a}}, Astrophys. J., 875, L1,
  \dodoi{10.3847/2041-8213/ab0ec7}

\bibitem[{Akiyama {et~al.}(2019{\natexlab{b}})}]{Akiyama:2019bqs}
---. 2019{\natexlab{b}}, Astrophys. J., 875, L4,
  \dodoi{10.3847/2041-8213/ab0e85}

\bibitem[{Akiyama {et~al.}(2019{\natexlab{c}})}]{Akiyama:2019fyp}
---. 2019{\natexlab{c}}, Astrophys. J., 875, L5,
  \dodoi{10.3847/2041-8213/ab0f43}

\bibitem[{Akiyama {et~al.}(2019{\natexlab{d}})}]{Akiyama:2019eap}
---. 2019{\natexlab{d}}, Astrophys. J., 875, L6,
  \dodoi{10.3847/2041-8213/ab1141}

\bibitem[{Amarilla \& Eiroa(2012)}]{Amarilla:2011fx}
Amarilla, L., \& Eiroa, E.~F. 2012, Phys. Rev., D85, 064019,
  \dodoi{10.1103/PhysRevD.85.064019}

\bibitem[{Amarilla \& Eiroa(2013)}]{Amarilla:2013sj}
---. 2013, Phys. Rev., D87, 044057, \dodoi{10.1103/PhysRevD.87.044057}

\bibitem[{Amarilla {et~al.}(2010)Amarilla, Eiroa, \& Giribet}]{Amarilla:2010zq}
Amarilla, L., Eiroa, E.~F., \& Giribet, G. 2010, Phys. Rev., D81, 124045,
  \dodoi{10.1103/PhysRevD.81.124045}

\bibitem[{Amir \& Ghosh(2016)}]{Amir:2016cen}
Amir, M., \& Ghosh, S.~G. 2016, Phys. Rev., D94, 024054,
  \dodoi{10.1103/PhysRevD.94.024054}

\bibitem[{Amir {et~al.}(2018)Amir, Singh, \& Ghosh}]{Amir:2017slq}
Amir, M., Singh, B.~P., \& Ghosh, S.~G. 2018, Eur. Phys. J., C78, 399,
  \dodoi{10.1140/epjc/s10052-018-5872-3}

\bibitem[{Atamurotov {et~al.}(2013)Atamurotov, Abdujabbarov, \&
  Ahmedov}]{Atamurotov:2013sca}
Atamurotov, F., Abdujabbarov, A., \& Ahmedov, B. 2013, Phys. Rev., D88, 064004,
  \dodoi{10.1103/PhysRevD.88.064004}

\bibitem[{Ayon-Beato \& Garcia(1999)}]{AyonBeato:1999rg}
Ayon-Beato, E., \& Garcia, A. 1999, Phys. Lett., B464, 25,
  \dodoi{10.1016/S0370-2693(99)01038-2}

\bibitem[{Bambi(2013)}]{Bambi:2013sha}
Bambi, C. 2013, JCAP, 1308, 055, \dodoi{10.1088/1475-7516/2013/08/055}

\bibitem[{Bambi(2018)}]{Bambi:2017iyh}
---. 2018, Annalen Phys., 530, 1700430, \dodoi{10.1002/andp.201700430}

\bibitem[{Bambi \& Freese(2009)}]{Bambi:2008jg}
Bambi, C., \& Freese, K. 2009, Phys. Rev., D79, 043002,
  \dodoi{10.1103/PhysRevD.79.043002}

\bibitem[{Bambi \& Modesto(2013)}]{Bambi:2013ufa}
Bambi, C., \& Modesto, L. 2013, Phys. Lett., B721, 329,
  \dodoi{10.1016/j.physletb.2013.03.025}

\bibitem[{Bardeen(1968)}]{Bardeen:1968}
Bardeen, J. 1968, Tbilisi, USSR

\bibitem[{Bardeen(1973)}]{bardeen1973}
---. 1973, Black Holes, edited by C. DeWitt and BS DeWitt,  Gordon and Breach,
  New York

\bibitem[{Berti {et~al.}(2015)}]{Berti:2015itd}
Berti, E., {et~al.} 2015, Class. Quant. Grav., 32, 243001,
  \dodoi{10.1088/0264-9381/32/24/243001}

\bibitem[{Breton {et~al.}(2019)Breton, Lämmerzahl, \&
  Macías}]{Breton:2019arv}
Breton, N., Lämmerzahl, C., \& Macías, A. 2019, Class. Quant. Grav., 36,
  235022, \dodoi{10.1088/1361-6382/ab5169}

\bibitem[{Brinkerink {et~al.}(2019)}]{Brinkerink:2018bhw}
Brinkerink, C.~D., {et~al.} 2019, Astron. Astrophys., 621, A119,
  \dodoi{10.1051/0004-6361/201834148}

\bibitem[{Broderick {et~al.}(2009)Broderick, Fish, Doeleman, \&
  Loeb}]{Broderick:2008sp}
Broderick, A.~E., Fish, V.~L., Doeleman, S.~S., \& Loeb, A. 2009, Astrophys.
  J., 697, 45, \dodoi{10.1088/0004-637X/697/1/45}

\bibitem[{Broderick {et~al.}(2014)Broderick, Johannsen, Loeb, \&
  Psaltis}]{Loeb:2013lfa}
Broderick, A.~E., Johannsen, T., Loeb, A., \& Psaltis, D. 2014, Astrophys. J.,
  784, 7, \dodoi{10.1088/0004-637X/784/1/7}

\bibitem[{Broderick \& Narayan(2006)}]{Broderick:2005xa}
Broderick, A.~E., \& Narayan, R. 2006, Astrophys. J., 638, L21,
  \dodoi{10.1086/500930}

\bibitem[{Cardoso {et~al.}(2009)Cardoso, Miranda, Berti, Witek, \&
  Zanchin}]{Cardoso:2008bp}
Cardoso, V., Miranda, A.~S., Berti, E., Witek, H., \& Zanchin, V.~T. 2009,
  Phys. Rev., D79, 064016, \dodoi{10.1103/PhysRevD.79.064016}

\bibitem[{Carter(1968)}]{Carter:1968rr}
Carter, B. 1968, Phys. Rev., 174, 1559, \dodoi{10.1103/PhysRev.174.1559}

\bibitem[{Casares \& Jonker(2014)}]{Casares:2013tpa}
Casares, J., \& Jonker, P.~G. 2014, Space Sci. Rev., 183, 223,
  \dodoi{10.1007/s11214-013-0030-6}

\bibitem[{Chandrasekhar(1985)}]{Chandrasekhar:1985kt}
Chandrasekhar, S. 1985, {The mathematical theory of black holes} (Oxford:
  Oxford Univ. Press)

\bibitem[{Cunha \& Herdeiro(2018)}]{Cunha:2018acu}
Cunha, P. V.~P., \& Herdeiro, C. A.~R. 2018, Gen. Rel. Grav., 50, 42,
  \dodoi{10.1007/s10714-018-2361-9}

\bibitem[{Cunha {et~al.}(2015)Cunha, Herdeiro, Radu, \&
  Runarsson}]{Cunha:2015yba}
Cunha, P. V.~P., Herdeiro, C. A.~R., Radu, E., \& Runarsson, H.~F. 2015, Phys.
  Rev. Lett., 115, 211102, \dodoi{10.1103/PhysRevLett.115.211102}

\bibitem[{De~Vries(2000)}]{de2000}
De~Vries, A. 2000, Class. Quant. Grav., 17, 123

\bibitem[{Doeleman {et~al.}(2008)}]{Doeleman:2008qh}
Doeleman, S., {et~al.} 2008, Nature, 455, 78, \dodoi{10.1038/nature07245}

\bibitem[{Doeleman {et~al.}(2012)}]{Doeleman:2012zc}
Doeleman, S.~S., {et~al.} 2012, Science, 338, 355,
  \dodoi{10.1126/science.1224768}

\bibitem[{Eiroa \& Sendra(2018)}]{Eiroa:2017uuq}
Eiroa, E.~F., \& Sendra, C.~M. 2018, Eur. Phys. J., C78, 91,
  \dodoi{10.1140/epjc/s10052-018-5586-6}

\bibitem[{Fabian {et~al.}(1989)Fabian, Rees, Stella, \& White}]{Fabian:1989ej}
Fabian, A.~C., Rees, M.~J., Stella, L., \& White, N.~E. 1989, Mon. Not. Roy.
  Astron. Soc., 238, 729

\bibitem[{Falcke \& Markoff(2013)}]{Falcke:2013ola}
Falcke, H., \& Markoff, S.~B. 2013, Class. Quant. Grav., 30, 244003,
  \dodoi{10.1088/0264-9381/30/24/244003}

\bibitem[{Falcke {et~al.}(2000)Falcke, Melia, \& Agol}]{Falcke:1999pj}
Falcke, H., Melia, F., \& Agol, E. 2000, Astrophys. J., 528, L13,
  \dodoi{10.1086/312423}

\bibitem[{Fish {et~al.}(2014)Fish, Johnson, Lu, Doeleman, Bouman, Zoran,
  Freeman, Psaltis, Narayan, Pankratius, {et~al.}}]{fish2014imaging}
Fish, V.~L., Johnson, M.~D., Lu, R.-S., {et~al.} 2014, The Astrophysical
  Journal, 795, 134

\bibitem[{Gebhardt {et~al.}(2011)Gebhardt, Adams, Richstone, Lauer, Faber,
  Gultekin, Murphy, \& Tremaine}]{Gebhardt:2011yw}
Gebhardt, K., Adams, J., Richstone, D., {et~al.} 2011, Astrophys. J., 729, 119,
  \dodoi{10.1088/0004-637X/729/2/119}

\bibitem[{Gebhardt {et~al.}(2000)}]{Gebhardt:2000fk}
Gebhardt, K., {et~al.} 2000, Astrophys. J., 539, L13, \dodoi{10.1086/312840}

\bibitem[{Ghez {et~al.}(2008)}]{Ghez:2008ms}
Ghez, A.~M., {et~al.} 2008, Astrophys. J., 689, 1044, \dodoi{10.1086/592738}

\bibitem[{Ghosh(2015)}]{Ghosh:2014pba}
Ghosh, S.~G. 2015, Eur. Phys. J., C75, 532,
  \dodoi{10.1140/epjc/s10052-015-3740-y}

\bibitem[{Giddings \& Psaltis(2018)}]{Giddings:2016btb}
Giddings, S.~B., \& Psaltis, D. 2018, Phys. Rev., D97, 084035,
  \dodoi{10.1103/PhysRevD.97.084035}

\bibitem[{Gillessen {et~al.}(2009)Gillessen, Eisenhauer, Trippe, Alexander,
  Genzel, Martins, \& Ott}]{Gillessen:2008qv}
Gillessen, S., Eisenhauer, F., Trippe, S., {et~al.} 2009, Astrophys. J., 692,
  1075, \dodoi{10.1088/0004-637X/692/2/1075}

\bibitem[{Grenzebach {et~al.}(2014)Grenzebach, Perlick, \&
  Lämmerzahl}]{Grenzebach:2014fha}
Grenzebach, A., Perlick, V., \& Lämmerzahl, C. 2014, Phys. Rev., D89, 124004,
  \dodoi{10.1103/PhysRevD.89.124004}

\bibitem[{Grenzebach {et~al.}(2015)Grenzebach, Perlick, \&
  Lämmerzahl}]{Grenzebach:2015oea}
---. 2015, Int. J. Mod. Phys., D24, 1542024, \dodoi{10.1142/S0218271815420249}

\bibitem[{H{\"a}ring \& Rix(2004)}]{haring2004black}
H{\"a}ring, N., \& Rix, H.-W. 2004, The Astrophysical Journal Letters, 604, L89

\bibitem[{Held {et~al.}(2019)Held, Gold, \& Eichhorn}]{Held:2019xde}
Held, A., Gold, R., \& Eichhorn, A. 2019, JCAP, 1906, 029,
  \dodoi{10.1088/1475-7516/2019/06/029}

\bibitem[{Hioki \& Maeda(2009)}]{Hioki:2009na}
Hioki, K., \& Maeda, K.-i. 2009, Phys. Rev., D80, 024042,
  \dodoi{10.1103/PhysRevD.80.024042}

\bibitem[{Hod(2009)}]{Hod:2009td}
Hod, S. 2009, Phys. Rev., D80, 064004, \dodoi{10.1103/PhysRevD.80.064004}

\bibitem[{Johannsen(2013{\natexlab{a}})}]{Johannsen:2013rqa}
Johannsen, T. 2013{\natexlab{a}}, Phys. Rev., D87, 124017,
  \dodoi{10.1103/PhysRevD.87.124017}

\bibitem[{Johannsen(2013{\natexlab{b}})}]{Johannsen:2015qca}
---. 2013{\natexlab{b}}, Astrophys. J., 777, 170,
  \dodoi{10.1088/0004-637X/777/2/170}

\bibitem[{Johannsen(2016)}]{Johannsen:2016uoh}
---. 2016, Class. Quant. Grav., 33, 124001,
  \dodoi{10.1088/0264-9381/33/12/124001}

\bibitem[{Johannsen \& Psaltis(2010)}]{Johannsen:2010ru}
Johannsen, T., \& Psaltis, D. 2010, Astrophys. J., 718, 446,
  \dodoi{10.1088/0004-637X/718/1/446}

\bibitem[{Johannsen \& Psaltis(2011)}]{Johannsen:2011dh}
---. 2011, Phys. Rev., D83, 124015, \dodoi{10.1103/PhysRevD.83.124015}

\bibitem[{Kerr(1963)}]{Kerr:1963ud}
Kerr, R.~P. 1963, Phys. Rev. Lett., 11, 237, \dodoi{10.1103/PhysRevLett.11.237}

\bibitem[{Konoplya {et~al.}(2016)Konoplya, Rezzolla, \&
  Zhidenko}]{Konoplya:2016jvv}
Konoplya, R., Rezzolla, L., \& Zhidenko, A. 2016, Phys. Rev., D93, 064015,
  \dodoi{10.1103/PhysRevD.93.064015}

\bibitem[{Konoplya \& Stuchlik(2017)}]{Konoplya:2017wot}
Konoplya, R.~A., \& Stuchlik, Z. 2017, Phys. Lett., B771, 597,
  \dodoi{10.1016/j.physletb.2017.06.015}

\bibitem[{Konoplya \& Zhidenko(2019)}]{Konoplya:2019goy}
Konoplya, R.~A., \& Zhidenko, A. 2019, Phys. Rev., D100, 044015,
  \dodoi{10.1103/PhysRevD.100.044015}

\bibitem[{Kumar {et~al.}(2019)Kumar, Ghosh, \& Wang}]{Kumar:2019pjp}
Kumar, R., Ghosh, S.~G., \& Wang, A. 2019, Phys. Rev., D100, 124024,
  \dodoi{10.1103/PhysRevD.100.124024}

\bibitem[{Long {et~al.}(2019)Long, Wang, Chen, \& Jing}]{Long:2019nox}
Long, F., Wang, J., Chen, S., \& Jing, J. 2019, JHEP, 10, 269,
  \dodoi{10.1007/JHEP10(2019)269}

\bibitem[{Luminet(1979)}]{Luminet:1979nyg}
Luminet, J.~P. 1979, Astron. Astrophys., 75, 228

\bibitem[{Matt \& Perola(1992)}]{matt1992iron}
Matt, G., \& Perola, G.~C. 1992, Mon. Not. Roy. Astron. Soc., 259, 433

\bibitem[{McClintock {et~al.}(2014)McClintock, Narayan, \&
  Steiner}]{McClintock:2013vwa}
McClintock, J.~E., Narayan, R., \& Steiner, J.~F. 2014, Space Sci. Rev., 183,
  295, \dodoi{10.1007/s11214-013-0003-9}

\bibitem[{McClintock {et~al.}(2011)McClintock, Narayan, Davis, Gou, Kulkarni,
  Orosz, Penna, Remillard, \& Steiner}]{McClintock:2011zq}
McClintock, J.~E., Narayan, R., Davis, S.~W., {et~al.} 2011, Class. Quant.
  Grav., 28, 114009, \dodoi{10.1088/0264-9381/28/11/114009}

\bibitem[{Melia \& Falcke(2001)}]{Melia:2001dy}
Melia, F., \& Falcke, H. 2001, Ann. Rev. Astron. Astrophys., 39, 309,
  \dodoi{10.1146/annurev.astro.39.1.309}

\bibitem[{Mizuno {et~al.}(2018)Mizuno, Younsi, Fromm, Porth, De~Laurentis,
  Olivares, Falcke, Kramer, \& Rezzolla}]{Mizuno:2018lxz}
Mizuno, Y., Younsi, Z., Fromm, C.~M., {et~al.} 2018, Nat. Astron., 2, 585,
  \dodoi{10.1038/s41550-018-0449-5}

\bibitem[{Narayan(2005)}]{Narayan_2005}
Narayan, R. 2005, New Journal of Physics, 7, 199,
  \dodoi{10.1088/1367-2630/7/1/199}

\bibitem[{Narayan \& McClintock(2012)}]{Narayan:2011eb}
Narayan, R., \& McClintock, J.~E. 2012, Mon. Not. Roy. Astron. Soc., 419, L69,
  \dodoi{10.1111/j.1745-3933.2011.01181.x}

\bibitem[{Narayan {et~al.}(2008)Narayan, McClintock, \&
  Shafee}]{Narayan:2007ks}
Narayan, R., McClintock, J.~E., \& Shafee, R. 2008, AIP Conf. Proc., 968, 265,
  \dodoi{10.1063/1.2840411}

\bibitem[{Newman {et~al.}(1965)Newman, Couch, Chinnapared, Exton, Prakash, \&
  Torrence}]{Newman:1965my}
Newman, E.~T., Couch, R., Chinnapared, K., {et~al.} 1965, J. Math. Phys., 6,
  918, \dodoi{10.1063/1.1704351}

\bibitem[{Papnoi {et~al.}(2014)Papnoi, Atamurotov, Ghosh, \&
  Ahmedov}]{Papnoi:2014aaa}
Papnoi, U., Atamurotov, F., Ghosh, S.~G., \& Ahmedov, B. 2014, Phys. Rev., D90,
  024073, \dodoi{10.1103/PhysRevD.90.024073}

\bibitem[{Perlick {et~al.}(2018)Perlick, Tsupko, \&
  Bisnovatyi-Kogan}]{Perlick:2018iye}
Perlick, V., Tsupko, O.~{\relax Yu}., \& Bisnovatyi-Kogan, G.~S. 2018, Phys.
  Rev., D97, 104062, \dodoi{10.1103/PhysRevD.97.104062}

\bibitem[{Psaltis {et~al.}(2008)Psaltis, Perrodin, Dienes, \&
  Mocioiu}]{Psaltis:2007cw}
Psaltis, D., Perrodin, D., Dienes, K.~R., \& Mocioiu, I. 2008, Phys. Rev.
  Lett., 100, 091101, \dodoi{10.1103/PhysRevLett.100.091101,
  10.1103/PhysRevLett.100.119902}

\bibitem[{Reid {et~al.}(2014)}]{Reid:2014boa}
Reid, M.~J., {et~al.} 2014, Astrophys. J., 783, 130,
  \dodoi{10.1088/0004-637X/783/2/130}

\bibitem[{Rezzolla \& Zhidenko(2014)}]{Rezzolla:2014mua}
Rezzolla, L., \& Zhidenko, A. 2014, Phys. Rev., D90, 084009,
  \dodoi{10.1103/PhysRevD.90.084009}

\bibitem[{Schee \& Stuchlik(2009)}]{Schee:2008kz}
Schee, J., \& Stuchlik, Z. 2009, Int. J. Mod. Phys., D18, 983,
  \dodoi{10.1142/S0218271809014881}

\bibitem[{Schodel {et~al.}(2002)}]{Schodel:2002vg}
Schodel, R., {et~al.} 2002, Nature, 419, 694, \dodoi{10.1038/nature01121}

\bibitem[{Shafee {et~al.}(2006)Shafee, McClintock, Narayan, Davis, Li, \&
  Remillard}]{Shafee:2005ef}
Shafee, R., McClintock, J.~E., Narayan, R., {et~al.} 2006, Astrophys. J., 636,
  L113, \dodoi{10.1086/498938}

\bibitem[{Shen {et~al.}(2005)Shen, Lo, Liang, Ho, \& Zhao}]{Shen:2005cw}
Shen, Z.-Q., Lo, K.~Y., Liang, M.~C., Ho, P. T.~P., \& Zhao, J.~H. 2005,
  Nature, 438, 62, \dodoi{10.1038/nature04205}

\bibitem[{Simpson \& Visser(2020)}]{simpson2020regular}
Simpson, A., \& Visser, M. 2020, Universe, 6, 8

\bibitem[{Singh \& Ghosh(2018)}]{Singh:2017vfr}
Singh, B.~P., \& Ghosh, S.~G. 2018, Annals Phys., 395, 127,
  \dodoi{10.1016/j.aop.2018.05.010}

\bibitem[{Stefanov {et~al.}(2010)Stefanov, Yazadjiev, \&
  Gyulchev}]{Stefanov:2010xz}
Stefanov, I.~Z., Yazadjiev, S.~S., \& Gyulchev, G.~G. 2010, Phys. Rev. Lett.,
  104, 251103, \dodoi{10.1103/PhysRevLett.104.251103}

\bibitem[{Steiner {et~al.}(2009)Steiner, McClintock, Remillard, Narayan, \&
  Gou}]{Steiner:2009af}
Steiner, J.~F., McClintock, J.~E., Remillard, R.~A., Narayan, R., \& Gou, L.
  2009, Astrophys. J., 701, L83, \dodoi{10.1088/0004-637X/701/2/L83}

\bibitem[{Steiner {et~al.}(2011)Steiner, Reis, McClintock, Narayan, Remillard,
  Orosz, Gou, Fabian, \& Torres}]{Steiner:2010bt}
Steiner, J.~F., Reis, R.~C., McClintock, J.~E., {et~al.} 2011, Mon. Not. Roy.
  Astron. Soc., 416, 941, \dodoi{10.1111/j.1365-2966.2011.19089.x}

\bibitem[{Synge(1966)}]{Synge:1966}
Synge, J. 1966, Mont. Not. R. Astron. Soc., 131, 463

\bibitem[{Takahashi(2004)}]{Takahashi:2004xh}
Takahashi, R. 2004, J. Korean Phys. Soc., 45, S1808, \dodoi{10.1086/422403}

\bibitem[{Tsukamoto {et~al.}(2014)Tsukamoto, Li, \& Bambi}]{Tsukamoto:2014tja}
Tsukamoto, N., Li, Z., \& Bambi, C. 2014, JCAP, 1406, 043,
  \dodoi{10.1088/1475-7516/2014/06/043}

\bibitem[{Tsupko(2017)}]{Tsupko:2017rdo}
Tsupko, O.~{\relax Yu}. 2017, Phys. Rev., D95, 104058,
  \dodoi{10.1103/PhysRevD.95.104058}

\bibitem[{Vagnozzi \& Visinelli(2019)}]{Vagnozzi:2019apd}
Vagnozzi, S., \& Visinelli, L. 2019, Phys. Rev., D100, 024020,
  \dodoi{10.1103/PhysRevD.100.024020}

\bibitem[{Walsh {et~al.}(2013)Walsh, Barth, Ho, \& Sarzi}]{Walsh:2013uua}
Walsh, J.~L., Barth, A.~J., Ho, L.~C., \& Sarzi, M. 2013, Astrophys. J., 770,
  86, \dodoi{10.1088/0004-637X/770/2/86}

\bibitem[{Wang {et~al.}(2019)Wang, Xu, \& Wei}]{Wang:2018prk}
Wang, H.-M., Xu, Y.-M., \& Wei, S.-W. 2019, JCAP, 1903, 046,
  \dodoi{10.1088/1475-7516/2019/03/046}

\bibitem[{Wang {et~al.}(2017)Wang, Chen, \& Jing}]{Wang:2017hjl}
Wang, M., Chen, S., \& Jing, J. 2017, JCAP, 1710, 051,
  \dodoi{10.1088/1475-7516/2017/10/051}

\bibitem[{Wang {et~al.}(2018)Wang, Chen, \& Jing}]{Wang:2018eui}
---. 2018, Phys. Rev., D98, 104040, \dodoi{10.1103/PhysRevD.98.104040}

\bibitem[{Wilkins(1972)}]{wilkins1972bound}
Wilkins, D.~C. 1972, Phys. Rev., D5, 814

\bibitem[{Yan(2019)}]{Yan:2019etp}
Yan, H. 2019, Phys. Rev., D99, 084050, \dodoi{10.1103/PhysRevD.99.084050}

\bibitem[{Young(1976)}]{Young:1976zz}
Young, P.~J. 1976, Phys. Rev., D14, 3281, \dodoi{10.1103/PhysRevD.14.3281}

\bibitem[{Younsi {et~al.}(2016)Younsi, Zhidenko, Rezzolla, Konoplya, \&
  Mizuno}]{Younsi:2016azx}
Younsi, Z., Zhidenko, A., Rezzolla, L., Konoplya, R., \& Mizuno, Y. 2016, Phys.
  Rev., D94, 084025, \dodoi{10.1103/PhysRevD.94.084025}

\end{thebibliography}
\bibliographystyle{aasjournal}
\end{document}